\documentclass[11pt,preprint]{aastex}

\newcommand{\ul}{\underline{\hspace{40pt}}}
\newcommand{\beq}{\begin{equation}}
\newcommand{\eeq}{\end{equation}}
\newcommand{\beqarr}{\begin{eqnarray}}
\newcommand{\eeqarr}{\end{eqnarray}}
\newcommand{\bml}{\begin{mathletters}}
\newcommand{\eml}{\end{mathletters}}

\newcommand{\cc}{{\rm{cm}^{-3}}}
\newcommand{\kms}{{\rm{km~s^{-1}}}}

\newcommand{\kB}{k_{\rm B}}

\newcommand{\Msun}{M_{\odot}}

\newcommand{\zhat}{\mbox{\boldmath$\hat{z}$}}
\newcommand{\xhat}{\mbox{\boldmath$\hat{x}$}}
\newcommand{\yhat}{\mbox{\boldmath$\hat{y}$}}
\newcommand{\nhat}{\mbox{\boldmath$\hat{n}$}}

\newcommand{\Bvec}{\mbox{\boldmath $B$}}

\newcommand{\vvec}{\mbox{\boldmath $v$}}
\newcommand{\jvec}{\mbox{\boldmath $j$}}
\newcommand{\yvec}{\mbox{\boldmath $y$}}
\newcommand{\Fvec}{\mbox{\boldmath $F$}}

\newcommand{\Htwo}{{\rm H}_{2}}
\newcommand{\mn}{m_{\rm n}}
\newcommand{\mi}{m_{\rm i}}
\newcommand{\nn}{n_{\rm n}}
\newcommand{\nno}{n_{\rm{n,0}}}
\newcommand{\rhon}{\rho_{\rm n}}
\newcommand{\rhoni}{\rho_{\rm{n,0}}}
\newcommand{\sign}{\sigma_{\rm n}}
\newcommand{\signi}{\sigma_{\rm{n,0}}}
\newcommand{\Zi}{Z_{0}}

\newcommand{\threehalf}{\mbox{$\frac{3}{2}$}}

\newcommand{\sigin}{\langle \sigma w \rangle_{\rm{i\Htwo}}}

\newcommand{\mui}{\mu_{0}}

\newcommand{\nni}{n_{\rm i}}

\newcommand{\vnx}{v_{{\rm n},x}}
\newcommand{\vny}{v_{{\rm n},y}}
\newcommand{\vix}{v_{{\rm i},x}}
\newcommand{\viy}{v_{{\rm i},y}}

\newcommand{\gx}{g_{x}}
\newcommand{\gy}{g_{y}}

\newcommand{\kx}{k_{x}}
\newcommand{\ky}{k_{y}}
\newcommand{\kz}{k_{z}}

\newcommand{\FT}{{\cal F}}

\newcommand{\tni}{\tau_{\rm{ni}}}
\newcommand{\tnii}{\tau_{\rm{ni,0}}}
\newcommand{\tniitil}{\tilde{\tau}_{\rm{ni,0}}}

\newcommand{\Pext}{P_{\rm{ext}}}
\newcommand{\Pexttil}{\tilde{P}_{\rm{ext}}}

\newcommand{\Ceff}{C_{\rm{eff}}}
\newcommand{\Cefftil}{\tilde{C}_{\rm{eff}}}
\newcommand{\Ceffitil}{\tilde{C}_{\rm{eff},0}}

\newcommand{\Beq}{B_{z,\rm{eq}}}

\newcommand{\Bref}{B_{\rm ref}}
\newcommand{\Breftil}{\tilde{B}_{\rm ref}}
\newcommand{\BxZ}{B_{{}_{x,Z}}}
\newcommand{\ByZ}{B_{{}_{y,Z}}}

\newcommand{\Fmagx}{F_{{\rm mag},x}}
\newcommand{\Fmagy}{F_{{\rm mag},y}}

\newcommand{\lamt}{\lambda_{\rm T}}
\newcommand{\lamtm}{\lambda_{\rm T,m}}
\newcommand{\lamms}{\lambda_{\rm MS}}
\newcommand{\lammsm}{\lambda_{\rm MS,m}}
\newcommand{\lammin}{\lambda_{\rm g,m}}

\newcommand{\Alf}{\mbox{Alfv\'{e}n}}

\newcommand{\delsign}{\delta\sign}
\newcommand{\delvnx}{\delta\vnx}
\newcommand{\delvny}{\delta\vny}
\newcommand{\delBeq}{\delta\Beq}
\newcommand{\delsigamp}{\delta\sigma_{\rm n,a}}

\newcommand{\tgrow}{\tau_{\rm g}}
\newcommand{\vani}{\tilde{V}_{\rm{A,0}}}
\newcommand{\vmsi}{\tilde{V}_{\rm{MS,0}}}

\begin{document}

\title{Formation and Collapse of Nonaxisymmetric Protostellar
Cores in Planar Magnetic Interstellar Clouds: Formulation
of the Problem and Linear Analysis}

\shorttitle{Nonaxisymmetric Collapse in Magnetic Clouds I.}
\shortauthors{G. E. Ciolek \& S. Basu}
\slugcomment{To appear in {\it The Astrophysical Journal}}

\author{Glenn E. Ciolek\altaffilmark{1,2}
and Shantanu Basu\altaffilmark{3}}

\altaffiltext{1}{New York Center for Studies on the Origin of Life
(NSCORT).} 
\altaffiltext{2}{Department of Physics, Applied Physics, and Astronomy,
Rensselaer Polytechnic Institute, 110 8th Street, Troy, NY 12180;
cioleg@rpi.edu.}
\altaffiltext{3}{Department of Physics and Astronomy, University of
Western Ontario, London, Ontario N6A~3K7, Canada; basu@astro.uwo.ca.}

\begin{abstract}
We formulate the problem of the formation and collapse of 
nonaxisymmetric protostellar cores in weakly ionized, self-gravitating,
magnetic molecular clouds. In our formulation, molecular clouds are 
approximated as isothermal, thin (but with finite thickness) sheets.
We present the governing dynamical equations for the multifluid system
of neutral gas and ions, including ambipolar diffusion, and also a 
self-consistent treatment of thermal pressure, gravitational, and 
magnetic (pressure and tension) forces. The dimensionless free 
parameters characterizing model clouds are discussed. The response of 
cloud models to linear perturbations is also examined, with 
particular emphasis on length and time scales for the growth of 
gravitational instability in magnetically subcritical and supercritical
clouds. We investigate their dependence on a cloud's initial 
mass-to-magnetic-flux ratio $\mui$ (normalized to the critical
value for collapse), the dimensionless initial neutral-ion collision
time $\tniitil$, and also the relative external pressure exerted on
a model cloud $\Pexttil$. Among our results, we find that 
nearly-critical model clouds have significantly larger characteristic 
instability lengthscales than do more distinctly sub- or supercritical 
models. Another result is that the effect of a greater external pressure
is to reduce the critical lengthscale for instability. Numerical 
simulations showing the evolution of model clouds during the linear 
regime of evolution are also presented, and compared to the results of 
the dispersion analysis. They are found to be in agreement with the 
dispersion results, and confirm the dependence of the characteristic 
length and time scales on parameters such as $\mui$ and $\Pexttil$.
\end{abstract}

\section{Introduction}
\subsection{Background}
In recent years there has been a steady and significant accumulation 
of observational data that support the idea that magnetic fields play a
pivotal role in the evolution of interstellar molecular clouds and star
formation. For instance, polarization maps of extended regions in clouds
reveal large-scale magnetic fields that in several instances are aligned
with the short axis of clouds (e.g. Pereyra \& Magalh\~{a}es 2004), 
which would be consistent with support from magnetic forces 
perpendicular to field lines. Infrared, far-infrared, and submillimeter
maps of more localized regions show ordered field configurations with 
curvature, including ``hourglass" shapes, that suggest dynamic 
interaction of magnetic fields and molecular cloud gas (Schleuning 1998;
Schleuning et al. 2000; Fujiyoshi et al. 2001). 

Interferometric polarization maps of molecular cloud cores by Lai et al.
(2001, 2002) indicate similar ordered field structures. Furthermore, 
using a modification of the  Chandrasekhar-Fermi (CF) method 
(Chandrasekhar \& Fermi 1953), they estimated magnetic field strengths 
and energies that were large enough to exceed observed gas turbulence
in their survey targets. A modified CF method was applied to the 
sub-mm polarization measurements of several prestellar cores by Crutcher
et al. (2004), and they found that the mass-to-magnetic flux ratios of 
their cores were clustered about (within a factor $\sim 2$ above and 
below) the critical value for gravitational collapse. This result is 
consistent with earlier OH Zeeman determinations of the mass-to-flux 
ratios of protostellar cores (Crutcher 2004). Likewise, Curran et al.
(2004) obtained sub-mm polarimetric images of two high-mass protostellar
cores; a CF analysis of their data yielded mass-to-flux ratios for their
two objects that were also nearly equal to the critical value for 
collapse. Finally, comparing CF results with OH Zeeman measurements
for the immediate environment of the L184 prestellar core, Crutcher 
(2004) concludes that it may be an example of a nearly or
barely critical core contained within a magnetically subcritical (i.e.,
supported) envelope. The grand picture painted by the various 
observational data above is one in which magnetic fields to a large 
degree control the structure and dynamics of star-forming molecular 
clouds, and has many elements of the original, unified theory of 
magnetically regulated star formation put forth early on by 
Mouschovias (1976, 1977, 1978, 1979; see, also, the review by 
Mouschovias \& Ciolek 1999). 

Ambipolar diffusion --- the drift of neutral matter with respect to 
plasma and magnetic fields --- initiates protostellar core formation in
magnetically supported clouds. That is, because of imperfect collisional
coupling between neutrals and charged particles (including dust grains;
Ciolek \& Mouschovias 1993, 1994), gravitationally-driven diffusion of 
matter in the inner flux tubes of clouds redistributes mass and magnetic
flux that leads to the formation and subsequent dynamical collapse of 
supercritical cores or fragments within massive, subcritical envelopes.
This process was exhibited in the two-dimensional axisymmetric magnetic
cloud simulations of Fiedler \& Mouschovias (1993), and the axisymmetric
thin-disk model calculations by Ciolek \& Mouschovias (1994, 1995) and 
Basu \& Mouschovias (1994, 1995a,b). Uniformly, these studies 
demonstrated the formation of  supercritical protostellar cores embedded
in subcritical envelopes. They were also able to follow the resultant 
collapse of these cores well into the stage of dynamical infall, ending
at the time when the central density was enhanced by a factor 
$\sim 10^{6}$.

Ciolek \& Basu (2000; hereafter CB00) applied one of their axisymmetric
magnetic disk models of core formation by ambipolar diffusion to the 
L1544 starless core. This model provided a good theoretical fit to the 
extended, subsonic infall profile as well as the distribution of mass 
in that core found by the earlier, complementary studies of 
Tafalla et al.  (1998) and Williams et al. (1999). The predicted 
magnetic field strength in the CB00 model was later found to be in 
agreement with OH Zeeman observations of L1544 (Crutcher \& Troland 
2000; CB00). An analysis of submillimeter and millimeter
maps and theoretical models of the dust temperature distribution in 
L1544 were found to be consistent with the emission distribution
calculated from the CB00 model (Zucconi, Walmsley, \& Galli
2001). Subsequent detailed, multispectral line maps used to further 
examine the kinematics within the L1544 core have also shown reasonable 
agreement between measured infall speeds and the CB00 predictions 
(Caselli et al. 2002). Finally, a multispectral survey by 
Crapsi et al. (2004) has found that some of the infall features of the 
prestellar core L1521F (which, like L1544, is in the Taurus 
complex) are comparable to those of the CB00 model.

However, despite the apparent successes of the aforementioned ambipolar 
diffusion models in describing the earliest stages of protostellar 
collapse, they are inherently limited because of their underlying
assumption of axisymmetry. Axisymmetry is clearly a highly restrictive
and idealized assumption, and is not likely to naturally occur 
in molecular clouds. In fact, sub-mm maps of star-forming clouds 
frequently indicate distinctively inhomogeneous and irregular 
large-scale structure and multiple cores (e.g., Andr\'{e}, Motte, 
\& Belloche 2001; Motte, Andr\'{e}, \& Neri 1998). Investigations
at these wavelengths reveal irregularities on the scale of individual 
cores as well (e.g., Bacmann et al. 2000). Moreover, the observational 
studies of the L1544 core cited above find that it is definitely not
axisymmetric. This is not a surprising result, as statistical analyses 
of data sets and catalogs of cores (e.g., Jones, Basu, \& Dubinski 2001;
Jones \& Basu 2002; Kerton et al.  2003) reveal that their shapes can 
generally be best fit with a distribution of triaxial ellipsoids, which
are typically more oblate than prolate.

Indebetouw \& Zweibel (2000) simulated the formation of supercritical
cores by ambipolar diffusion in infinitesimally thin, magnetic layers. 
Their models were two-dimensional and nonaxisymmetric, and they focused
primarily on clouds that were initially subcritical. They followed the
evolution of cores to when the maximum column density was a factor 
$\lesssim 5$ above the initial mean column density, and the irregular 
cores that developed were reminiscent of the sub-mm observations cited 
above.\footnote{Some of the results in the interiors of their cores 
(scaling of magnetic field strength with central density, subsonic 
infall speeds, etc.) were similar to those found in the earlier 
axisymmetric model simulations over the same range of central density.
This is also true for the nonaxisymmetric critical cloud model of Basu
\& Ciolek (2004 --- see the denser, inner core regions in their Fig. 1).
This agreement is probably the reason why the axisymmetric CB00 model is
a reasonable physical model for the interior of the asymmetric L1544 
prestellar core.}
They also showed that the cores in their calculations developed on a 
timescale corresponding to the maximum linear growth rate of 
gravitational instability, and had a size equal to the corresponding 
wavelength (Zweibel 1998; see, also, \S~3 below). Basu \& Ciolek (2004;
hereafter BC04) presented numerical models displaying the gravitational
collapse of multiple dense, asymmetric protostellar cores in thin (yet, 
not infinitesimally so) planar magnetic clouds. One of their model 
clouds had an initial reference state with a mass-to-magnetic flux 
ratio that was exactly the critical value ($\mui=1$, see \S\S~2.2 and 
3.1 below). Another model of theirs was initially supercritical by a 
factor of 2 (i.e., $\mui = 2$). They found that, in the initially 
critical model, transfer of mass by ambipolar diffusion resulted in the 
eventual formation of supercritical cores contained within subcritical 
envelopes, similar to that which occurred in their earlier axisymmetric
models that started with initially subcritical conditions. By contrast, 
the initially supercritical model of BC04 had a much more rapid and 
dynamical infall, with significantly greater infall speeds 
(sonic and supersonic) extending over much larger regions. Based on the
physically distinct and different predictions of these two models, they 
noted that molecular clouds that are supercritical by more than a 
factor $\sim 2$ would be incompatible with the generally observed
subsonic infall motions.
\vspace{-2ex}
\subsection{Outline}
In this paper, we present the formulation for modeling the formation 
and nonaxisymmetric collapse of protostellar cores in planar 
magnetic clouds. In \S~2 we describe the fundamental assumptions and 
derive the necessary system of governing equations for a model cloud.
The equations are put in nondimensional form, and the resulting free 
parameters of a model and their physical meaning are described. Our 
numerical method of solving the governing equations is also discussed.
To provide a basis for understanding the underlying physics of ambipolar
diffusion and gravitational collapse in magnetic sheets, the stability 
of model clouds is examined in \S~3 by linearizing and Fourier-analyzing
the governing equations. The results of the stability analysis are also
compared with full numerical simulations of models in the limit of 
small-amplitude perturbations. In section \S~4 we summarize our results,
and discuss their relevance to forthcoming fully nonlinear studies of 
nonaxisymmetric core formation and collapse. 
\vspace{-2ex}
\section{Physical Formulation}
We model clouds as isothermal thin sheets with temperature $T$, 
embedded in a hot and tenuous external medium of constant pressure 
$\Pext$. This simplifying assumption, the {\em thin sheet} 
approximation, is based on the numerical simulations of 
ambipolar-diffusion-initiated formation of cores by Fiedler 
\& Mouschovias (1993), who found that two-dimensional, axially 
symmetric magnetically supported model molecular clouds rapidly flatten
and establish force-balance {\em along} the direction of 
magnetic field lines, even while contraction and dynamical collapse took
place in the direction {\em perpendicular} to the field. These 
results were soon afterward incorporated in the protostellar core 
formation studies of Ciolek \& Mouschovias (1993; hereafter CM93) and
Basu \& Mouschovias (1994; hereafter BM94), who modeled molecular clouds
as thin axisymmetric disks threaded by a vertical magnetic field, with 
hydrostatic equilibrium maintained along field lines at all times. Here
we again adopt the formulation of CM93 and BM94, but now forego the 
restriction of axisymmetry. 

The appealing feature of the thin-sheet approximation is that it can be
used to model the physical processes necessary to core and star 
formation in a theoretically tractable and realistic way, including a 
self-consistent treatment of both magnetic pressure and tension forces.
Another advantage of thin-sheet models is that because the physical 
equations have been simplified by integration in one dimension (here, in
the direction of the vertical magnetic field), the problem has been 
substantially reduced from its full three-dimensional complexity. 
Because of this, sheet models can provide significantly higher spatial 
resolution, and require much less computational time and storage than 
fully three-dimensional cloud models. There is also some observational 
justification for the use of the thin sheet approximation. The often 
observed alignment, or at least close correlation, between the projected
magnetic field direction from polarization measurements and the core 
minor axis is evidence for flattening along the field (Basu 2000). 
Furthermore, an analysis of core shapes by Jones et al. (2001) and 
Jones \& Basu (2002) implies that they are preferentially flattened 
along one direction.

Of course, sheet models may not account for all of the observed 
morphological features of molecular clouds and their envelopes. 
Especially for those with large internal velocity dispersion, which can
provide substantial support against self-gravity, and therefore be 
extended along magnetic field lines. It turns out though, that even in 
clouds or complexes with such substantial velocity dispersion or 
turbulence, thin-sheet models may still be used, so long as the 
application is restricted to dense sub-regions of the cloud. This has 
been demonstrated by Kudoh \& Basu (2003, 2006), who analyzed the 
nonlinear support of stratified molecular clouds due to an ensemble of 
driven hydromagnetic waves. They found that because of density 
stratification, the largest, supersonic, velocities are in the low 
density envelope of a cloud, while a dense region near the midplane 
has transonic or subsonic motions. Hence, in this situation, thin-sheet
models may reasonably be applied to a high-density fragment or 
sub-cloud, which will most likely be the site of subsequent star 
formation.  

The most significant systematic source of possible inaccuracy of a 
thin-sheet model is that it typically overestimates the strength of the 
gravitational field in the equatorial plane (i.e., the plane of the
disk) when compared to less condensed or concentrated systems. For 
instance, the critical wavelength for gravitational instability in very 
thin non-magnetic clouds is found to be half the value of the critical 
length found for an equilibrium layer with an exponential atmosphere.
This can be seen by comparing our equation (\ref{lamtdefeq}) derived in
\S~3.1, in the limit of zero external pressure, to equations (13-36) and
(13-42) of Spitzer (1978). However, the equilibrium layer calculation 
assumes an infinite vertical extent, which reduces the system's total 
gravitational binding energy; this is therefore an upper limit to the 
possible overestimate of calculating the gravitational field by using a 
thin disk. A better estimate was made by Kiguchi et al. (1987, see their
Fig. 11), who showed that the magnitude of the planar gravitational 
field for an infinitesimally thin disk exceeds that of a flattened 
cloud of finite extent by at most only $\sim 40\%$, with the largest 
discrepancy between the two occurring for a very small region around the
cloud center. BM94 presented a method of correcting for finite-thickness
effects in the magnetic thin-disk approximation, deriving a quadratic 
correction (in terms of the local disk thickness) to the gravitational 
field. For the density regimes we intend to study 
($\lesssim 10^{7}~\cc$), the BM94 correction had the effect of altering
the evolution of a model's core by about 10\% during the early 
stages of core evolution, to $\lesssim 30\%$ at much later stages.
Finally, we note that because the planar magnetic field in the sheet is
calculated the same way that the planar gravitational field is (see eqs.
[\ref{gravffteqa}] - [\ref{gyeqa}] and [\ref{magffteqa}] - [\ref{Byeqa}]
below), there is also a comparable overestimate in the magnitude of the
magnetic tension force. Since the magnetic tension force opposes
the gravitational force, the overestimate in the value of the
gravitational field due to the thin-sheet approximation is offset by a 
corresponding overestimate of the planar magnetic field, thereby 
reducing the net effect of overestimating both fields.

Thin-sheet models have been used by many workers in the field of 
molecular clouds and protostars. They were used early on by Narita, 
Hayashi, \& Miyama (1984) to study star formation in axisymmetric, 
non-magnetic clouds. As mentioned above, the thin-sheet approximation 
was developed and applied to ambipolar diffusion and core formation 
in isothermal magnetic molecular clouds by CM93 (who included
the effects of dust) and BM94 (who included rotation and magnetic 
braking). It was also later adopted by Li \& Shu (1997a,b), who studied
the equilibria and self-similar gravitational collapse of clouds with
frozen-in magnetic flux (i.e., no diffusion). The approach to
self-similar collapse during the later stage of core collapse with 
ambipolar diffusion in thin-disk clouds was described by Basu (1997).
Ciolek \& K\"{o}nigl (1998) incorporated the effect of the formation of
a central gravitating point mass (a protostar) in numerical
simulations of ambipolar-diffusion-driven dynamical core collapse in 
axisymmetric thin-disk clouds; a self-similar solution to this same
problem was also provided by Contopoulos, Ciolek, \& K\"{o}nigl (1998).
Tassis \& Mouschovias (2005a,b) also investigated this particular topic,
further extending the analysis by including a detailed calculation of 
multi-fluid effects on the conductivity of the infalling gas during the
later stages of supercritical core collapse and accretion.
 
A nonaxisymmetric `toy model' idealization of a magnetically critical,
flux-frozen turbulent sheet-like molecular cloud was suggested by
Allen \& Shu (2000). As noted in \S1.1, Indebetouw \& Zweibel (2000)
first presented nonaxisymmetric simulations of core formation by
ambipolar diffusion in infinitesimally thin magnetic sheets. 
BC04 presented models of nonaxisymmetric, gravitationally
collapsing cores in magnetically critical and supercritical 
finite-thickness sheet-like molecular clouds, including ambipolar 
diffusion and its consequent effect on the dynamical evolution of 
clouds and cores. Li \& Nakamura (2004) and Nakamura \& Li (2005) used 
the thin-sheet approximation to examine the combined effect of 
turbulent initial conditions and ambipolar diffusion in forming 
supercritical cores.

\vspace{-2ex}
\subsection{Fundamental Equations}
We present the necessary system of equations to model core formation in
weakly ionized, magnetic interstellar clouds. As stated above, our model
clouds are taken to be thin, with local vertical half-thickness 
$Z(x,y,t)$ in a Cartesian coordinate system. By a sheet being thin we 
mean that for any physical quantity $f(x,y,z,t)$, the criterion 
$f/|\nabla_{p}f| \gg Z$ is always satisfied, where 
$\nabla_{p} \equiv \xhat \partial /\partial x + \yhat \partial/\partial y$
is the planar gradient operator. The magnetic field threading a 
cloud is taken to have the form
\bml
\beqarr
\label{Beqna}
\Bvec(x,y,z,t)&=&\Beq(x,y,t)\zhat \hspace{16em} {\rm for}~ |z| \leq Z(x,y,t),~~~~\\
\label{Beqnb}
\Bvec(x,y,z,t)&=&B_{z}(x,y,z,t)\zhat + B_{x}(x,y,z,t)\xhat
+ B_{y}(x,y,z,t)\yhat \hspace{1.55em} {\rm for}~ |z| > Z(x,y,t),~~ 
\eeqarr
\eml
where $\Beq$ is the magnetic field strength in the equatorial plane
of the cloud. In the limit $|z| \rightarrow \infty$, 
$\Bvec \rightarrow \Bref \zhat$, where $\Bref$ is a constant, uniform 
reference magnetic field very far away from the sheet. From now on,
all physical quantities are understood to be a function of time $t$.  

Note that the condition on the planar gradient of physical quantities 
within the thin-sheet approximation implies that $Z$ is the lower limit 
to the scales that can be described by our model. (For the 
gravitationally unstable modes, this condition is always satisfied, as 
the critical lengths always exceed this value --- see eq. 
[\ref{lamtdefeq}].)

The unit normal 
vector to the upper and lower surfaces of the sheet is
\beq
\label{normvecteq}
\nhat(x,y, \pm Z) = \frac{\pm \zhat \mp \left[(\partial Z/\partial x)\xhat 
+ (\partial Z/\partial y)\yhat\right]}
{\left[1 + (\partial Z/\partial x)^{2} + (\partial Z/\partial y)^{2}\right]^{1/2}}~~.
\eeq
Use of the integral form of Gauss's law yields the continuity equation
for the normal component of the magnetic field across the upper and
lower surfaces of the sheet,
\beq
\label{gaussconteq}
B_{z}(x,y,\pm Z) - B_{x}(x,y,\pm Z)\frac{\partial Z}{\partial x}
- B_{y}(x,y,\pm Z)\frac{\partial Z}{\partial y} = \Beq(x,y) .
\eeq

The system of multifluid equations necessary to model a
weakly-ionized, magnetic, self-gravitating, isothermal molecular cloud 
is given by equations ($9a$)-($9m$) of CM93. (In this study we 
ignore the dynamical effect of interstellar dust grains. Hence, those 
terms representing grain contributions appearing in the system of fluid
equations in CM93 are neglected.) As was done in CM93 and BM94, we
simplify these basic equations in the thin-sheet approximation by 
vertically integrating them from $z=-Z(x,y)$ to $z=+Z(x,y)$. Doing so 
for the equation of mass continuity yields
\beq
\label{massconteqa}
\frac{\partial \sign}{\partial t} 
+ \frac{\partial}{\partial x} (\sign \vnx)
+ \frac{\partial}{\partial y} (\sign \vny) = 0~~,
\eeq
where $\sign(x,y) \equiv \int_{-Z}^{+Z} \rhon(x,y) dz$ is the
mass column density, and $\vnx$ and $\vny$ are respectively the $x$- 
and $y$-components of the neutral velocity. In deriving equation
(\ref{massconteqa}) we have used the chain rule to obtain the velocity
of the surface of the disk,
\beq
\label{Zveloceq}
\frac{d Z}{dt} = \vnx \frac{\partial Z}{\partial x} 
+ \vny \frac{\partial Z}{\partial y}~~.
\eeq 
We have also employed a ``one-zone" approximation,
where we assume $z$-independence of physical quantities such
as $\rhon$, and the planar velocity components $\vnx$, and $\vny$. 
This approximation is also used for the planar components of the 
gravitational acceleration $\gx$ and $\gy$ that appear in the equations
below.

To derive the $x$- and $y$-components of the force equation (per
unit area) for the neutrals, we use equation (\ref{normvecteq}) and the
total (thermal plus Maxwell) stress tensor
\beq
\label{stresstenseq}
{\sf T} = - \left(\rhon C^{2} + \frac{B^{2}}{8 \pi}\right){\sf 1} + 
\frac{\Bvec \Bvec}{4 \pi}~~,  
\eeq
where ${\sf 1}$ is the identity tensor, $C = (\kB T/\mn)^{1/2}$ is 
the isothermal speed of sound; $\kB$ is the Boltzmann constant, and
$\mn$ is the mean mass of a neutral particle (= 2.33 a.m.u. for an 
$\Htwo$ gas with a 10\% He abundance by number). Using equation
(\ref{gaussconteq}), the symmetry conditions on $\Beq$ and $B_{z}$
and the antisymmetry conditions on $B_{x}$ and $B_{y}$ about the
equatorial plane, along with the divergence theorem, the integrated 
force equations are 
\bml
\beqarr
\label{mtmxeqa}
\frac{\partial}{\partial t}\left(\sign \vnx\right)
+ \frac{\partial}{\partial x}\left(\sign \vnx^{2}\right) 
+ \frac{\partial}{\partial y}\left(\sign \vnx \vny\right)
&=& \sign \gx - \Ceff^{2} \frac{\partial \sign}{\partial x}
+ \Fmagx ~~,\\
\label{mtmyeqa}
\frac{\partial}{\partial t}\left(\sign \vny\right)
+ \frac{\partial}{\partial x}\left(\sign \vny \vnx\right)
+ \frac{\partial}{\partial y}\left(\sign \vny^{2}\right) 
&=& \sign \gy - \Ceff^{2} \frac{\partial \sign}{\partial y}
+ \Fmagy~~, 
\eeqarr
\eml
where 
\bml
\beqarr
\label{Fmagxeqa}
\Fmagx &=& \frac{\Beq}{2\pi}\left(\BxZ - 
Z \frac{\partial \Beq}{\partial x}\right) \nonumber \\
& &
+ \frac{1}{4 \pi} 
\frac{\partial Z}{\partial x}\left[\BxZ^{2} + \ByZ^{2}
+  2 \Beq 
\left(\BxZ \frac{\partial Z}{\partial x}+\ByZ \frac{\partial Z}{\partial y}\right)
+ \left(\BxZ \frac{\partial Z}{\partial x} + \ByZ \frac{\partial Z}{\partial y} \right)^{2} \right]~,
\hspace{2em}\\ 
\label{Fmagyeqa}
\Fmagy &=& \frac{\Beq}{2 \pi} \left(\ByZ 
- Z \frac{\partial \Beq}{\partial y}\right) \nonumber \\ 
& &
+ \frac{1}{4 \pi} 
\frac{\partial Z}{\partial y}\left[\BxZ^{2} + \ByZ^{2}
+  2 \Beq 
\left(\BxZ \frac{\partial Z}{\partial x}+\ByZ \frac{\partial Z}{\partial y}\right)
+ \left(\BxZ \frac{\partial Z}{\partial x} + \ByZ \frac{\partial Z}{\partial y} \right)^{2} \right]~,
\eeqarr
\eml
$\BxZ \equiv B_{x}(x,y,+Z)$, $\ByZ \equiv B_{y}(x,y,+Z)$,
and
\beq
\label{Ceffeqa}
\Ceff^{2} \equiv \frac{\pi}{2} G \sign^{2}
\frac{\left(3 \Pext + \frac{\pi}{2} G \sign^{2}\right)}
{\left(\Pext + \frac{\pi}{2} G \sign^{2}\right)^{2}}
C^{2}
\eeq
is the local effective sound speed. $G$ is the gravitational constant.
The second term on the right hand sides of equations (\ref{Fmagxeqa}) 
and (\ref{Fmagyeqa}) represent small modifications to the magnetic force
due to planar gradients of the half-thickness $Z$.

The equations for $\vix$ and $\viy$, the $x$- and $y$-components
of the ion velocity, are similarly obtained by vertical integration
of the force equation for the ions. They are, respectively,
\bml
\beqarr
\label{vixeqa}
\vix &=&\vnx + \frac{\tni}{\sign} \Fmagx~~, \\
\label{viyeqa}
\viy &=&\vny + \frac{\tni}{\sign} \Fmagy~~, 
\eeqarr
\eml
where the neutral-ion collision (momentum-exchange) time
\beq
\label{tnidefeq}
\tni = 1.4 \left(\frac{\mi + m_{{}_{\Htwo}}}{\mi}\right) 
\frac{1}{\nni \sigin} ~~.
\eeq
The quantity $\mi$ is the ion mass, which we take to be 25 a.m.u.,
the mass of the typical atomic ($\rm{Na}^{+}$, $\rm{Mg}^{+}$)
and molecular ($\rm{HCO}^{+}$) ion species in clouds; $\sigin$ is the 
neutral-ion collision rate, and is equal to
$1.69 \times 10^{-9}~{\rm{cm}}^{3}~{\rm s}^{-1}$ for 
$\Htwo - {\rm{HCO}}^{+}$ collisions (McDaniel \& Mason 1973). 
The factor of 1.4 in equation (\ref{tnidefeq}) accounts for the fact
that the inertia of helium is neglected in calculating the slowing-down
time of the neutrals by collisions with ions. (Further discussion
on this point can be found in \S~2.1 of Mouschovias \& Ciolek 1999.) 
For the ion number density we assume a power-law 
behavior of the form
\beq
\label{rhoieq}
\nni =  {\cal K} \left(\frac{\nn}{10^{5}~\cc}\right)^{k}~~,
\eeq
where ${\cal K}$ ($\simeq 3 \times 10^{-3}~\cc$) and $k$ ($\simeq 1/2$) 
are constants. In reality, the exponent $k$ is also a function of 
density (Ciolek \& Mouschovias 1998), due to the fact that ambipolar 
diffusion can deplete the abundance of dust grains in a contracting core
(Ciolek \& Mouschovias 1996), which alters the rate of capture and 
recombination of ions and electrons on grain surfaces. However, we 
ignore this effect in our models for the time being.

Integrating the $z$-component of the force equation for the neutrals
from $z=0$ to $z=+Z$, and requiring that there be hydrostatic 
equilibrium along field lines yields
\beq
\label{rhoneqa}
\rhon C^{2} = \frac{\pi}{2} G \sign^{2} + \Pext 
+ \frac{\left(\BxZ^{2} + \ByZ^{2} + \left[\BxZ\frac{\partial Z}{\partial x}
+ \ByZ \frac{\partial Z}{\partial y}\right]^{2}\right)}{8 \pi}~,
\eeq
where we have used the Gaussian relation for a thin sheet,
$g_{z}(x,y,+Z) = -2 \pi G \sign(x,y,+Z)$.
As discussed in CM93 and BM94, the first two terms on the right
side of equation (\ref{rhoneqa}) represent the self-gravitational stress
and external pressure acting on a sheet, respectively. The latter term 
--- not included in our earlier studies (e.g., see eq.  [26] of CM93) 
--- is the total magnetic ``pinching" term due to magnetic pressure and 
tension stresses that act to compress the sheet. Although this last term
is generally smaller than
the others, we retain it nevertheless in our models, since it costs very
little computationally to include it.

We use Poisson's equation for a very thin sheet to solve for the 
gravitational potential $\psi$:
\beq
\label{gravpoissoneq}
\nabla^{2} \psi(x, y, z) = 4 \pi G \sign(x, y) \delta(z) .
\eeq
Imposing the boundary condition 
$\lim_{|z| \rightarrow \infty} \psi(x, y, z) \rightarrow 0$, equation 
(\ref{gravpoissoneq}) can be solved by the method of Fourier transforms
(e.g., Wyld 1976; Byron \& Fuller 1992). Doing so, one finds for $z=0$,
\beq
\label{gravffteqa}
\FT[\psi(x,y,0)] = -2 \pi G \frac{\FT[\sign(x,y)]}{\kz} ,
\eeq
where ${\FT}[f]$ is the two-dimensional Fourier transform of the 
function $f$, and $\kz = (\kx^{2} + \ky^{2})^{\onehalf}$ is a function
of the planar wave numbers $\kx$ and $\ky$. Hence, at any time $t$ we 
can determine $\FT[\psi]$ by calculating $\FT[\sign]$. Inverting the 
transform then yields $\psi$ at that time, and from that we can 
calculate the gravitational field 
\bml
\beqarr
\label{gxeqa}
\gx(x,y,0) = -\frac{\partial \psi(x,y,0)}{\partial x} , \\ 
\label{gyeqa}
\gy(x,y,0) = -\frac{\partial \psi(x,y,0)}{\partial y} . 
\eeqarr
\eml

The magnetic field components $\BxZ$ and $\ByZ$ can be gotten in a
similar fashion. Above the sheet, the magnetic field can be written in
the form $\Bvec(x,y, z > +Z) = \Bvec^{\prime}(x,y,z) + \Bref\zhat$,
where $\Bvec^{\prime}$ is the reduced magnetic field. The assumption 
that the constant pressure external medium is hot and 
tenuous implies that it is also current-free ($\jvec_{\rm{ext}} = 0$).
Consequently, Ampere's law gives $\nabla \times \Bvec = 0$ in the region
above the sheet. Therefore, $\Bvec^{\prime} = -\nabla \Psi$, where 
$\Psi$ is a scalar magnetic potential, that, because of Gauss's law 
($\nabla \cdot \Bvec =0$), satisfies Laplace's equation, 
\beq
\label{laplaceeq}
\nabla^{2}\Psi(x,y,z > +Z) = 0.
\eeq 
From our earlier comments on the external field we have 
$\Bvec^{\prime} \rightarrow 0$ very far from the sheet. Therefore,
\beq
\label{farfieldconeq}
\lim_{z \rightarrow \infty} \Psi(x,y,z) = 0~~.
\eeq
The boundary condition on $\Psi$ at $z=+Z$ is derived from the 
continuity equation (\ref{gaussconteq}) for the normal 
component of the magnetic field across the top surface of the sheet,
which is, written in terms of $\Psi$,
\bml
\beq
\label{gausspoteqa}
\frac{\partial \Psi(x,y,+Z)}{\partial z}
-\frac{\partial \Psi(x,y,+Z)}{\partial x}\frac{\partial Z}{\partial x}
-\frac{\partial \Psi(x,y,+Z)}{\partial y}\frac{\partial Z}{\partial y}
= -\left[\Beq(x,y) - \Bref\right]~~.
\eeq
In the thin-sheet approximation, this becomes
\beq
\label{gausspoteqb}
\lim_{+Z \rightarrow 0} \frac{\partial \Psi(x,y,+Z)}{\partial z}
= -\left[\Beq(x,y) - \Bref\right]~~.
\eeq
\eml
The solution of the magnetic Laplace equation (\ref{laplaceeq}) for
$\Psi$, subject to the boundary conditions (\ref{farfieldconeq}) and
(\ref{gausspoteqb}), can also be performed by Fourier transform, in
analogy to what is done for the gravitational potential $\psi$.
In the limit $+Z \rightarrow 0$, we find
\beq
\label{magffteqa}
\FT[\Psi(x,y,0)] = \frac{\FT\left[\Beq(x,y) - \Bref\right]}{\kz}~~.
\eeq
Once $\Psi$ is obtained, by inverting the transform, it follows that
\bml
\beqarr
\label{Bxeqa}
\BxZ &=&\lim_{+Z \rightarrow 0} -\frac{\partial \Psi(x,y,+Z)}{\partial x},
\\
\label{Byeqa}
\ByZ &=&\lim_{+Z \rightarrow 0} -\frac{\partial \Psi(x,y,+Z)}{\partial y}~~.
\eeqarr
\eml

To close our system of equations, the evolution of the equatorial 
magnetic field in the plane of the sheet $\Beq(x,y)$ is governed by
the magnetic induction equation. For the density range we consider
in our cloud models, $10^{3}~\cc \lesssim \nn \lesssim 10^{8}~\cc$, 
the magnetic field is effectively ``frozen" in the ion-electron plasma 
(see \S\S~2.2 - 2.4 of Mouschovias \& Ciolek 1999 for a discussion). 
Hence, advection of magnetic flux in our system is described by 
\beq
\label{inducteqa}
\frac{\partial \Beq}{\partial t} =
-\frac{\partial}{\partial x}\left(\Beq \vix\right)
-\frac{\partial}{\partial y}\left(\Beq \viy\right) ~~. 
\eeq
\vspace{-4ex}
\subsection{Boundary Conditions, Uniform Background State and Initial 
Conditions}
As described in the preceding section, model clouds are assumed to be 
isothermal thin planar sheets of infinite extent. We follow the 
evolution in a square Cartesian region of size $L$, with 
$-L/2 \leq x \leq L/2$, $-L/2 \leq y \leq L/2$. Periodic boundary 
conditions are used for all physical quantities.

Model clouds are initially characterized by a static, uniform 
background state with constant column density $\signi$ and equatorial 
magnetic field $\Bref \zhat$ (i.e., $B_{z,{\rm{eq}},0} = \Bref$). From 
equations (\ref{mtmxeqa}), (\ref{mtmyeqa}), (\ref{gxeqa}), 
(\ref{gyeqa}), (\ref{Bxeqa}), and (\ref{Byeqa}), it is seen that all 
forces --- thermal, gravitational, and magnetic --- are identically 
equal to zero in the background state. Evolution is initiated in a cloud
by superposing a set of random column density perturbations,
$\delta\sign(x,y)$ ($\ll \signi$, typically) on the uniform background 
state at time $t=0$. To maintain the same local mass-to-flux ratio
$\sign/\Beq$ in the initial state as in the reference state, the 
magnetic field is simultaneously perturbed, with 
$\delta \Beq/\Bref = \delta \sign/\signi$. 
\vspace{-2ex}
\subsection{Dimensionless Equations and Free Parameters}
The actual system of equations we solve are dimensionless versions of
the equations presented in \S~2.1 above. We adopt the following
normalizations: the velocity unit is $[v]=C$, the column density
unit is $[\sigma]=\signi$, unit of acceleration
is $[a]= 2 \pi G \signi$, the time unit is $[t] = C/2\pi G \signi$,
and the length unit is $[l]= C^{2}/2 \pi G \signi$. From this system
one can also construct a unit of magnetic field strength,
$[B] = 2 \pi G^{1/2} \signi$. With these normalizations, the
dimensionless equations that govern the evolution of
a model cloud are
\bml
\beqarr
\label{massconteqb}
\frac{\partial \sign}{\partial t} &=& 
- \frac{\partial}{\partial x}(\sign \vnx)
- \frac{\partial}{\partial y}(\sign \vny) ~~, \\
\label{mtmxeqb}
\frac{\partial}{\partial t}\left(\sign \vnx\right) &=&
- \frac{\partial}{\partial x}\left(\sign \vnx^{2}\right)
- \frac{\partial}{\partial y}\left(\sign \vny \vnx\right) 
+\sign \gx - \Cefftil^{2} \frac{\partial \sign}{\partial x} 
+ \Fmagx~~,\\
\label{mtmyeqb}
\frac{\partial}{\partial t}\left(\sign \vny\right) &=&
- \frac{\partial}{\partial x}\left(\sign \vny \vnx\right)
- \frac{\partial}{\partial y}\left(\sign \vny^{2}\right) 
+\sign \gy - \Cefftil^{2} \frac{\partial \sign}{\partial y} 
+ \Fmagy~~,\\
\label{Fmagxeqb}
\Fmagx &=& \Beq \left(\BxZ - Z \frac{\partial \Beq}{\partial x}\right)
\nonumber \\
& & + \frac{1}{2}\frac{\partial Z}{\partial x}
\left[\BxZ^{2} + \ByZ^{2} 
+ 2 \Beq 
\left(\BxZ \frac{\partial Z}{\partial x} + \ByZ \frac{\partial Z}{\partial y}\right)
+ \left(\BxZ \frac{\partial Z}{\partial x} + \ByZ \frac{\partial Z}{\partial y}\right)^{2} \right],
\hspace{2.5em}\\ 
\label{Fmagyeqb}
\Fmagy &=& \Beq \left(\ByZ - Z \frac{\partial \Beq}{\partial y}\right)
\nonumber \\
& & + \frac{1}{2}\frac{\partial Z}{\partial y}
\left[\BxZ^{2} + \ByZ^{2} 
+ 2 \Beq 
\left(\BxZ \frac{\partial Z}{\partial x} + \ByZ \frac{\partial Z}{\partial y}\right)
+ \left(\BxZ \frac{\partial Z}{\partial x} + \ByZ \frac{\partial Z}{\partial y}\right)^{2} \right], 
\hspace{2.5em}\\ 
\label{Ceffeqb}
\Cefftil^{2} &=&  \sign^{2}\frac{\left(3 \Pexttil + \sign^{2}\right)}
{\left(\Pexttil + \sign^{2}\right)^{2}}~~, \\
\label{inducteqb}
\frac{\partial \Beq}{\partial t} &=&
-\frac{\partial}{\partial x}\left(\Beq \vix\right)
-\frac{\partial}{\partial y}\left(\Beq \viy\right) ~~, \\ 
\label{vixeqb}
\vix &=&\vnx + \frac{\tniitil}{\sign}\left(\frac{\rhon}{\rhoni}\right)^{k} \Fmagx~~,\\ 
\label{viyeqb}
\viy &=&\vny + \frac{\tniitil}{\sign}\left(\frac{\rhon}{\rhoni}\right)^{k} \Fmagy~~\\ 
\label{rhoneqb}
\rhon &=& \frac{1}{4} \left(\sign^{2} + \Pexttil 
+ \BxZ^{2} + \ByZ^{2} + \left[\BxZ\frac{\partial Z}{\partial x}
+ \ByZ \frac{\partial Z}{\partial y}\right]^{2}\right)~~, \\
\label{Zeq}
Z &=& \frac{\sign}{2 \rhon}~~, \\
\label{gravffteqb}
{\FT}[\psi] &=& -\frac{\FT[\sign]}{\kz}~~,\\
\label{gxeqb}
\gx &=& -\frac{\partial \psi}{\partial x}~~, \\ 
\label{gyeqb}
\gy &=& -\frac{\partial \psi}{\partial y}~~, \\ 
\label{magffteqb}
{\FT}[\Psi] &=& \frac{\FT[\Beq - \Breftil]}{\kz}~~,\\
\label{Bxeqb}
\BxZ &=& -\frac{\partial \Psi}{\partial x}~~, \\ 
\label{Byeqb}
\ByZ &=& -\frac{\partial \Psi}{\partial y}~~, 
\eeqarr
\eml
where $\rhoni = (1 + \Pexttil)/4$ is the dimensionless neutral mass 
density of the background state, and $\psi$ and $\Psi$ represent
their values in the equatorial plane.

The above equations are, for the most part, the Cartesian analogs
of the axisymmetric thin-disk equations presented in CM93 (their
[66a]-[66q]) and BM94 (their [34a]-[34m]). A similar set of non-ideal
MHD equations were used to model nonaxisymmetric core formation by 
Indebetouw \& Zweibel (2000), with one particular exception: they
modeled clouds as {\em infinitesimally} thin, with $Z=0$, and neglected
the effect of magnetic pressure (i.e., the terms dependent on
$Z$ in the equations above, and also in CM93 and BM94). This means 
that the stabilizing effect of magnetosound modes were not accounted 
for in the models of Indebetouw \& Zweibel. Although this may be valid 
in some cases, there are, as we show below, some regions in the 
physically allowed parameter space for clouds in which magnetic pressure
cannot be considered negligible compared to magnetic tension. The 
omission of magnetic pressure in these instances leads to inaccuracies 
in the length- and timescales for the onset of gravitational instability
in magnetic clouds.

Equations (\ref{massconteqb})-(\ref{Byeqb}) have several
non-dimensional parameters. $\Pexttil \equiv 2 \Pext/\pi G\signi^{2}$
is the ratio of the external pressure acting on the sheet to
the vertical self-gravitational stress of the background state. 
The effect of ambipolar diffusion is expressed by the dimensionless 
initial neutral-ion collision time,
$\tniitil \equiv 2 \pi G \signi \tnii/C$. The limit 
$\tniitil \rightarrow \infty$ corresponds to extremely poor neutral-ion 
collisional coupling, so that the ions and magnetic field have 
no effect on the neutrals. In the opposite limit,
$\tniitil =0$, the neutrals are perfectly coupled to the ions, due
to frequent collisions, and the magnetic field will be essentially 
frozen in the neutral matter. The parameter $k$ (=1/2, typically) is 
the exponent in the power-law expression (\ref{rhoieq}) used to 
calculate the ion density as a function of the neutral density. As we 
noted earlier, this constant power-law assumption is only an 
approximation, since ambipolar diffusion has been shown to make $k$ a
function of $\nn$ in models with a more realistic ion chemistry network
(Ciolek \& Mouschovias 1998). Finally,
$\Breftil = \Bref/2 \pi G^{1/2} \signi$ is the dimensionless
magnetic field strength of the background state. For the units we have 
chosen for the column density and the magnetic field, the dimensionless
mass-to-magnetic-flux ratio of the background state is
\beq
\label{muieq}
\mui \equiv 2 \pi G^{{}^{\onehalf}} \frac{\signi}{\Bref} = \frac{1}{\Breftil}~~. 
\eeq
This also happens to be the mass-to-flux ratio in units of the critical
value for gravitational collapse, $1/2 \pi G^{\onehalf}$ (Nakano \&
Nakamura 1978; see, also, \S~3.1 below). Models with $\mui < 1$
($\Breftil > 1$) are subcritical clouds, while those with $\mui > 1$
($\Breftil < 1$) are supercritical. 

Normally, we set $\signi$ by specifying the temperature $T$ and the 
density $\nno$ of the background state. Using equation (\ref{rhoneqa}), 
and the fact that $C =0.188(T/10~{\rm K})^{1/2}~\kms$ for 
$\mn = 2.33$ a.m.u.,
\beq
\label{signivaleq}
\signi = \frac{3.63 \times 10^{-3}}{(1 + \Pexttil)^{\onehalf}} 
\left(\frac{\nno}{10^{3}~\cc}\right)^{\onehalf}
\left(\frac{T}{10~{\rm K}}\right)^{\onehalf}~~{\rm g}~{\rm{cm}^{-2}}~~.
\eeq
Both $C$ and $\signi$ are normalizing units that we have 
adopted for our cloud models. The units for length and time have the 
scalings
\beq
\label{lengthuniteq}
[l] = 7.48 \times 10^{-2} \left(\frac{T}{10 {\rm K}}\right)^{\onehalf}
\left(\frac{10^{3}~\cc}{\nno}\right)^{\onehalf}
\left(1+ \Pexttil\right)^{\onehalf}
~~{\rm{pc}},
\eeq
and
\beq
\label{timeuniteq}
[t]= 3.98 \times 10^{5}
\left(\frac{10^{3}~\cc}{\nno}\right)^{\onehalf}
\left(1 + \Pexttil \right)^{\onehalf}
~~{\rm{yr}}.
\eeq
The unit of mass is then
\beq
\label{massuniteq}
[M] = 9.76 \times 10^{-2} \left(\frac{T}{10~{\rm K}}\right)^{\threehalf}
\left(\frac{10^{3}~\cc}{\nno}\right)^{\onehalf}\left(1 + \Pexttil\right)^{\onehalf}
~\Msun~.
\eeq

It follows from equations (\ref{tnidefeq}), (\ref{rhoieq}), and 
(\ref{signivaleq}) that
\beq
\label{tniitilvaleq}
\tniitil = \frac{0.241}{(1 + \Pexttil)^{\onehalf}}
\left(\frac{3 \times 10^{-3} \cc}{{\cal K}}\right)
\left(\frac{10^{5}~\cc}{\nno}\right)^{k-\onehalf}~~.
\eeq
From equations (\ref{muieq}) and (\ref{signivaleq}), a model's
reference magnetic field is given by
\beq
\label{Brefvaleq}
\Bref = \frac{5.89 \times 10^{-6}}{\mui (1 + \Pexttil)^{\onehalf}}
\left(\frac{\nno}{10^{3}~\cc}\right)^{\onehalf}
\left(\frac{T}{10~{\rm K}}\right)^{\onehalf} ~~{\rm G}. 
\eeq
\vspace{-2ex}
\subsection{Numerical Method of Solution}
The system of dimensionless partial differential equations 
presented in the preceding section is solved by the method of lines
(e.g., Schiesser 1991). This was used
in the axisymmetric models of Morton, Mouschovias, \& Ciolek (1994), 
Ciolek \& Mouschovias (1994, 1995), and Basu \& Mouschovias (1994, 
1995a, b), and the nonaxisymmetric models presented in BC04. In this 
method, spatial derivatives within the PDEs are 
approximated by finite differences. The square computational domain of 
size $L \times L $ is discretized by dividing the region into $N^{2}$ 
uniform cells of size $L/N \times L/N$. $L$ is typically chosen to be
a factor $\sim 4$ greater than the wavelength of maximum gravitational 
instability $\lamtm$ (see \S~3.1 below). Three-point centered 
differences are used to approximate gradients. Advection of mass and 
magnetic flux is performed with the monotonic upwind scheme of van Leer
(1979). The spatial discretization converts the system of PDEs to a 
system of coupled ordinary differential equations of the form 
$d{\yvec}/dt = \Fvec (\yvec,t)$, where $d\yvec/dt$, $\yvec$, and $\Fvec$
are all vectors of dimension $V N^{2}$ ($V$ is the number of dependent 
variables). An implicit Adams-Bashforth-Moulton method is then used to 
time-integrate the resulting system of ODEs. 

At each time step, numerical solution of the Fourier transform and also
the inverse transform of quantities in the equations for the 
gravitational and magnetic potentials ([\ref{gravffteqb}] and 
[\ref{magffteqb}]) is carried out with standard two-dimensional Fast 
Fourier Transform (FFT) techniques (e.g., Press et al. 1996).  

\section{Stability of Cloud Models}
We now consider the response of model clouds to small-amplitude
disturbances or perturbations. Such an analysis will illuminate the 
basic physics of gravitational instability in weakly ionized, 
sheet-like magnetic clouds, and will also provide the basis for 
understanding the length and time scales over which instabilities will 
develop in fully nonlinear calculations. 

The linear stability of isothermal self-gravitating equilibrium layers 
with frozen-in magnetic fields was examined early on by Nakano 
\& Nakamura (1978). They determined the critical mass-to-flux ratio for
such objects, which is discussed further below. The gravitational 
stability of weakly-ionized and thin (but with finite thickness) 
axisymmetric magnetic disks was investigated by Morton (1991)
and Morton \& Mouschovias (1991, unpublished). Their analysis is very 
similar to the one that we present in this paper, and we reprise many of
their results in the following section. A later, independent linear 
study of partially-ionized magnetic disks/sheets was also presented by 
Zweibel (1998). Similar to that which was done in Indebetouw \& Zweibel
(2000), Zweibel (1998) studied clouds that were infinitesimally thin
($Z=0$), and concentrated on those that are magnetically subcritical.
\vspace{-2ex}
\subsection{Linearization and Analysis}
As is frequently done, the unperturbed zero-order
(background) state of a model is assumed to be uniform and static. The
dimensionless equations presented in \S~2.3 above are linearized to 
first-order by writing, for any physical quantity,
\beqarr
\label{lineardefeq}
f(x,y,t) &=& f_{0} + \delta f \nonumber \\
&=& f_{0} + \delta f_{{}_{\rm a}}e^{i(\kx x + \ky y - \omega t)} ,
\eeqarr
where $f_{0}$ refers to the unperturbed state, $\delta f$ is the
perturbation and $\delta f_{{}_{\rm a}}$ is its amplitude, $\kx$ and 
$\ky$ are again the $x$ and $y$ wave numbers, and $\omega$ is the 
complex angular frequency. The perturbations are small, with 
$|\delta f_{{}_{\rm a}}| \ll f_{0}$. With regard to velocities, which
have $\vvec_{0} = 0$ because of the static assumption, it is understood
that $|\delta\vvec| \ll$ characteristic signal speeds of the 
multifluid system (i.e., sound speed, \Alf\ speed, etc.). 

For the assumed type of perturbation (\ref{lineardefeq}), we can set 
$\partial/\partial t \rightarrow -i \omega$, 
$\partial/\partial x \rightarrow i \kx$, and 
$\partial/\partial y \rightarrow i \ky$. 
The dimensionless equations for a model cloud then become, collecting
and retaining terms to first-order,
\bml
\beqarr 
\label{linconteq}
\omega~\delsign &=& \kx \delvnx + \ky \delvny~~, \\
\label{linvnxeq}
\omega~\delvnx &=& \frac{\kx}{\kz}\left(\Ceffitil^{2}\kz - 1\right)\delsign
+ \Breftil \frac{\kx}{\kz} \left(1 + \kz \Zi \right)\delBeq~~, \\ 
\label{linvnyeq}
\omega~\delvny &=& \frac{\ky}{\kz}\left(\Ceffitil^{2}\kz - 1\right)\delsign
+ \Breftil \frac{\ky}{\kz}\left(1 + \kz \Zi \right)\delBeq~~, \\ 
\label{linBeq}
\omega~\delBeq &=& \Breftil \kx \delvnx + \Breftil \ky \delvny
- i \tniitil \Breftil^{2} \kz\left(1 + \kz \Zi\right)\delBeq~~.
\eeqarr
\eml
From equation (\ref{Ceffeqb}), the reference state effective isothermal 
speed of sound is
\beq
\label{Ceffitildefeq}
\Ceffitil = \frac{\left(1 + 3\Pexttil\right)^{\onehalf}}{1 + \Pexttil}~.
\eeq
In deriving the linearized system (\ref{linconteq})-(\ref{linBeq}), 
we have used equations (\ref{gravffteqb})-(\ref{gyeqb}) to relate
gravitational field perturbations to column density perturbations,
(\ref{magffteqb})-(\ref{Byeqb}) to relate planar magnetic field 
perturbations to those of the equatorial vertical magnetic field,
and (\ref{vixeqb}) and (\ref{viyeqb}) to substitute the perturbed ion 
velocity components with those of the neutrals. The wavenumber $\kz$ has
the same meaning as previously defined. A mode is unstable if the 
imaginary part of the complex frequency
$\omega_{{}_{\rm I}} > 0$. The growth timescale of the instability is 
$\tgrow = 1/\omega_{{}_{\rm I}}$.

The fundamental physics of the linear system is readily discerned from
the various terms in equations (\ref{linconteq})-(\ref{linBeq}): 
thermal-pressure and self-gravitational forces are proportional to 
perturbations in the column density in equations (\ref{linvnxeq}) and 
(\ref{linvnyeq}), and magnetic tension  and pressure forces are 
proportional to perturbations in the equatorial magnetic field. The 
drift or diffusion of magnetic field and plasma with respect to the 
neutrals is represented by the term in the magnetic induction equation 
(\ref{linBeq}) that contains $\tniitil$; comparison with the linearized
mass continuity equation (\ref{linconteq}) shows that, in the limit 
$\tniitil = 0$, collisional coupling of the neutrals and ions (and 
therefore, the magnetic field) is instantaneous and perfect, and they 
all move together as a single fluid. In the opposite extreme, 
$\tniitil \gg 1$, neutral-ion collisions are infrequent, and the neutral
and ion-magnetic field fluids are increasingly decoupled and move 
independently of one another.

The solution of the full dispersion relation for the gravitationally 
unstable mode of the linearized equations (\ref{linconteq}) 
- (\ref{linBeq}) are presented in Figure 1 for (a) $\tniitil = 0$, 
(b) $\tniitil = 0.04$, (c) $\tniitil = 0.1$, (d) $\tniitil = 0.2$,
(e) $\tniitil = 1$, and (f) $\tniitil = 10$. The external pressure
parameter for these models is $\Pexttil = 0.1$, which sets the 
dimensionless effective sound speed $\Ceffitil = 1.04$. 
(Note that $\Ceffitil = 1$ at $\Pexttil = 0$ and 1. $\Ceffitil$ is 
maximal at $\Pexttil = 1/3$, and is equal to 1.061.) Displayed in each 
panel of Figure 1 is the growth time $\tgrow$ as a function of the 
wavelength $\lambda$ ($=2 \pi /\kz$), for
various values of $\mui = 1/\Breftil$. The separately labeled curves 
show the result for $\mui = 0.5$, 0.8, 1, 1.1, 2, and 10, respectively.
We are able to use $\lambda$ as the independent variable because the 
characteristic polynomial for our eigensystem is found to be only a 
function of $\kz = (\kx^{2} + \ky^{2})^{\onehalf}$. This means that to
this order of approximation, all perturbations are independent of the 
planar angle of propagation $\theta$ [$= \tan^{-1}(\ky/\kx)$].

Understanding the data presented in Figure $1a$ - $f$ is aided by 
examining the results for the instability growth timescale in the 
following two limits.
\begin{center}
{\it Limit 1: Flux-freezing, $\tniitil = 0$.}
\end{center}

In this limit, the neutrals, ions, and magnetic field respond as a 
single combined fluid, with the magnetic flux frozen in the neutrals,
and the resulting dispersion relation is found to be
\beq
\omega^{2} - \left(\Ceffitil^{2}  + \Breftil^{2} \Zi \right) \kz^{2}
- \left(\Breftil^{2} - 1 \right)\kz = 0. 
\eeq
The gravitationally unstable mode corresponds to one of the roots
of $\omega^{2} < 0$, and occurs for $\Breftil < 1$, or, equivalently, 
$\mui > 1$. The growth timescale for this mode can be written as
\beq
\label{tgrfluxfreeze}
\tgrow = \frac{\lambda}{\left[2 \pi (1 - \Breftil^{2})
(\lambda - \lamms)\right]^{\onehalf}}~~,
\eeq
for $\lambda \geq \lamms$, where 
\beq
\label{lammsdefeq}
\lamms \equiv 2 \pi \left(\frac{\Ceffitil^{2} + \Breftil^{2} \Zi}{1 - \Breftil^{2}}\right)
\eeq
is the critical or threshold wavelength for instability. 
The minimum growth timescale (= maximum growth rate) occurs at 
$\lammsm \equiv 2 \lamms$. The flux-freezing growth timescale 
(\ref{tgrfluxfreeze}) for our various model clouds is plotted as 
a function of $\lambda$ in Figure $1a$ - $f$, shown as open circles. 

We have given to the critical wavelength $\lamms$ the subscript `MS'
because it is the maximum lengthscale that can be supported by both 
magnetic and thermal pressure effects, and therefore is related to 
magnetosound modes. This can be readily seen by noting that the 
dimensionless \Alf\ speed in our model clouds is given by
\beq
\label{vanieq}
\vani = \frac{\Breftil}{\sqrt{2 \rhoni}} = \Breftil \Zi^{\onehalf} , 
\eeq
(see eq. [\ref{Zeq}]) and the fact that the dimensionless column density
is unity for an unperturbed cloud. It follows then that the isothermal
magnetosound speed (since we consider only isothermal perturbations
in our models) in the adopted set of units is 
\beqarr
\label{vmsieq}
\vmsi &=& \left(\Ceffitil^{2} + \vani^{2} \right)^{\onehalf} \nonumber \\
&=& \left(\Ceffitil^{2} + \Breftil^{2} \Zi \right)^{\onehalf} \nonumber \\
&=& \Ceffitil \left[1 + \frac{2}{\mui^{2}}\frac{(1 + \Pexttil)}{(1 + 3 \Pexttil)}\right]^{\onehalf}~.
\eeqarr
In the brackets of the last equality we have used eq. 
(\ref{Ceffitildefeq}) and the relation
\beq
\label{Ziniteq}
\Zi = \frac{2}{(1 + \Pexttil)}~~,
\eeq
which follows from the linearization of equations (\ref{rhoneqb}) and 
(\ref{Zeq}), to eliminate $\Ceffitil^{2}$ and $\Zi$ in terms of 
$\Pexttil$. Examination of equation (\ref{lammsdefeq}) reveals the 
presence of the magnetosound speed (\ref{vmsieq}) in this expression, 
thus identifying the combined action of thermal and magnetic pressure in
the support of a cloud against self-gravity. It also shows the 
importance of magnetic pressure in setting the instability timescale and
the wavelength of maximum growth rate. For clouds that are close to 
being critical ($\mui \sim 1$), neglecting this effect (e.g., Zweibel 
1998; Indebetouw \& Zweibel 2000) can significantly underestimate 
$\tgrow$ and $\lammsm$.

When there is negligible magnetic support, that is, when 
$\Breftil \rightarrow 0$ ($\mui \rightarrow \infty$) it follows that 
$\vmsi \rightarrow \Ceffitil$, and $\lamms \rightarrow \lamt$, where 
\beq
\label{lamtdefeq}
\lamt \equiv 2 \pi \Ceffitil^{2} 
= \pi \left(\frac{1 + 3 \Pexttil}{1 + \Pexttil}\right)\Zi
\eeq
is the critical thermal lengthscale. For this situation, the growth 
timescale for the unstable mode is still given by equation 
(\ref{tgrfluxfreeze}), with $\Breftil = 0$, and $\lamt$ replacing 
$\lamms$. The maximum growth rate for the unstable mode in this
circumstance then occurs at
$\lamtm \equiv 2 \lamt$.
\begin{center} 
{\it Limit 2: Stationary magnetic field lines, $\omega \delBeq = 0$.}
\end{center}

For this situation, which would be relevant to models with effective
ambipolar diffusion, and therefore, relatively weak coupling of the 
neutrals to the ions and magnetic field, the resulting dispersion 
relation is
\beq
\label{statdispeq}
\omega^{2} + \frac{i}{\tniitil} \omega 
- \left(\Ceffitil^{2} \kz^{2} - \kz \right) = 0.
\eeq
From this relation, one finds that an unstable mode exists for 
$\lambda > \lamt$, and has a growth timescale 
\beq
\label{tgrstationary}
\tgrow = \frac{ 2 \tniitil \lambda}
{\left[\lambda^{2} + 8 \pi \tniitil^{2}\left(\lambda - \lamt\right)\right]^{\onehalf} - \lambda}~~. 
\eeq
The minimum growth time for this mode is at the wavelength $\lamtm$, 
defined above. The growth time (\ref{tgrstationary}) is also 
displayed as a function of wavelength in Figure $1a$ - $f$ (crosses). 

As $\tniitil \rightarrow \infty$,  
$\tgrow \rightarrow \lambda/\sqrt{2\pi(\lambda - \lamt)}$, which is
identical to the result from equation (\ref{tgrfluxfreeze}) when 
$\Breftil = 0$. This is because in both of these circumstances the 
magnetic field does not affect the neutrals: 
$\tniitil \rightarrow \infty$ corresponds to when there is no 
collisional coupling between the neutrals and ions (and, hence, the 
magnetic field). The ions are completely ``invisible" to the neutrals 
in this situation, and there is no transmission of magnetic force to 
them via neutral-ion collisions. Similarly, when $\Breftil = 0$, there 
is no magnetic field, and therefore magnetic forces do not contribute to
the support or dynamics of a model cloud in that case.

With the results of the two limiting cases 1 and 2 described above
in hand, the underlying physics of the gravitationally unstable
modes presented in Figure $1a$ - $f$ is made more transparent. For
instance, the models with the magnetic field frozen into the neutrals
($\tniitil =0$) displayed in Figure $1a$ are seen to be in exact 
agreement with the growth timescale predicted by equation 
(\ref{tgrfluxfreeze}). Particularly noteworthy is the dependence on
the parameter $\mui$ ($= 1/\Breftil$) for these models: there is no 
unstable, gravitationally collapsing mode for $\mui < 1$. Moreover, 
models with $\mui \sim 1$ have minimum growth timescales at wavelengths
$\lammsm$ that are much larger than those for models with greater 
values of $\mui$. As an example, we note that $\lammsm$ for the model 
with $\mui = 1.1$ in Figure $1a$ is an order of magnitude greater than 
that in the model with $\mui = 2$. The actual value of the growth 
timescale for the $\mui = 1.1$ model is seen to be about an order of 
magnitude greater than that for the $\mui = 2$ model as well. This is a
consequence of there still being near-equality of gravitational and
magnetic forces when $\mui$ is very close to unity. 

When there is imperfect neutral-ion coupling ($\tniitil > 0$),
gravitational instability is possible even for 
model clouds with $\mui < 1$. This can be seen in panels $b$ - $f$ 
of Figure 1. The timescale for the instability is typically greater 
than that for the supercritical flux-frozen models (Fig. $1a$), but it 
is finite. For these models, a Jeans-like growth of density 
perturbations occurs with the collapse moderated by the retarding 
collisional forces exerted on the neutrals as they diffuse through the 
plasma and field. A result of this type was first noted by Langer
(1978). Models that are more subcritical are better approximated by the
stationary field limit described above. This
can be seen by the excellent agreement of the limiting growth timescale
given by equation (\ref{tgrstationary}) --- and also the wavelength of 
maximum growth ($\lamtm$) --- with the dispersion curve for the 
$\mui = 0.5$ model displayed in Figure $1b$ - $f$. As $\mui$ approaches
unity, there is a transition of the growth timescale behavior 
from the stationary field/ambipolar diffusion limit 
(eq. [\ref{tgrstationary}]) and toward the flux-freezing limit 
(eq. [\ref{tgrfluxfreeze}]). For $\mui$ close to unity, the instability
proceeds through a hybrid mode in which both ion-neutral drift and 
gravitational contraction with field-line dragging are active. As 
discussed above, the two limiting approximations approach one another
for models with $\tniitil \gg 1$ and $\mui \gg 1$, since either limit
describes clouds with effectively no magnetic support. This accounts for
the near equality of all the model curves seen in Figure $1f$. 

Figure 2 presents the wavelength $\lammin$, which is defined as the 
wavelength $\lambda$ that has the minimum growth time, as a 
function of $\mui$ for models with the same values of $\tniitil$ and 
$\Pexttil$ as in Figure $1a$ - $f$. For comparison, the dashed line in 
Figure 2 is the constant value of $\lamtm$ for those particular models. 
Consistent with our discussion above, we note that there is singular and
limiting behavior of this lengthscale for the model cloud with 
$\tniitil = 0$ and $\mui = 1$. The value of $\lammin$ is also seen to be
especially sensitive to the value of $\mui$ for near-critical clouds 
with $\tniitil > 0$, exhibiting a sharp, resonant-like peak in the 
region $\mui \sim 1$. Table 1 lists the ratio $\lammin/\lamtm$ at the 
peak for each of the non-flux-frozen models in Figure 2. The peak ratio
is largest for models with $\tniitil \lesssim 1$, while, for models with
$\tniitil \gg 1$, magnetic field effects are barely transmitted to the 
neutrals (because neutral-ion collisions are very rare), and this ratio
instead approaches unity. 

Based on these results, we expect then, that clouds that are marginally
or slightly supercritical will form cores that evolve more slowly, and 
have size scales (radii, and core spacing) {\em significantly greater} 
than that which will occur in clouds that are more highly supercritical.
In addition, nearly critical clouds will also have size scales that are
markedly larger than those in distinctly subcritical clouds. As we shall
see in a following nonlinear study, this specific dependence of the 
gravitational instability on the initial mass-to-flux ratio leads to 
notable differences in the physical characteristics of collapsing cores 
in magnetic interstellar clouds that should be easily discerned by 
observations. 

Figure $3a$ - $f$ shows the growth timescales for the same model clouds
as in Figure 1, but with the dimensionless external pressure 
$\Pexttil = 10$. In the limit of large $\Pexttil$, equation 
(\ref{rhoneqb}) shows that the cloud is pressure confined, with 
$Z \propto \sign$ (eq. [\ref{Zeq}]). As a result, a local peak in
$\sign$ is a peak in $Z$. For this situation then, the top surface of 
the cloud looks like a dome and the external pressure force has 
horizontal components that point inward to the dome's peak, acting in 
the same directions as the gravitational field components $\gx$ and 
$\gy$. This further enables the gravitational clumping, and decreases 
the growth timescale, appearing as a reduction of $\Ceffitil$ in our 
equations. Because of this, the dimensionless sound speed $\Ceffitil$ is
decreased by a factor of 2.05 from its value in the models with 
$\Pexttil = 0.1$. This also reduces the growth timescale $\tgrow$ and 
associated lengthscales $\lamms$ and $\lamt$, since they are also 
functions of $\Ceffitil$, and therefore, of $\Pexttil$. The net effect 
is to shift and reduce the instability timescale and critical 
wavelengths by a corresponding factor from those seen in the models with
lower external pressure. This means that clouds in regions with much 
greater external pressure (perhaps due to being embedded in a massive 
cloud complex or fragment, or adjoining hotter environments such as an 
HII region and/or shocked gas) will have characteristic sizes that are 
smaller when compared to clouds surrounded by lesser external pressures.
However, the general behavior and physics of the models as a function of
$\tniitil$ and $\mui$, including the applicability of the limiting 
analytical approximations discussed above, is similar to that seen in 
the previously described models with smaller $\Pexttil$. 

The wavelength with the minimum growth time $\lammin$ is shown as a 
function of $\mui$ in Figure 4 for models with the same values of 
$\tniitil$ and $\Pexttil$ as in Figure 3. It is again the case that, for
clouds that are almost critical and with $\tniitil \lesssim 1$, the 
wavelength that has the most rapid gravitational response is 
significantly greater than in models that are farther away from 
$\mui = 1$. The peak ratio values of $\lammin$ to $\lamtm$ for the 
models with $\tniitil > 0$ are also listed in Table 1. As before, this 
ratio is much greater for the models with $\tniitil \lesssim 1$.  
\vspace{-2ex}
\subsection{Comparison to numerical simulations}
We now compare the results of the dispersion analysis of the
preceding section to simulations of the early evolution of model clouds
governed by the full system of non-ideal magnetohydrodynamic equations
(\ref{massconteqb}) - (\ref{Byeqb}). This system of equations is solved
by a numerical code using the techniques described in \S~2.4. The code 
was previously used to generate results for two model clouds whose 
nonlinear evolution was presented by BC04. During the 
early evolution of a model cloud, physical variables do not change very
much from their initial state values. Hence, during this period the 
physical evolution should be close to that determined by the linear 
analysis above. Comparing the results of the early-time evolution of the
full simulations to those of the dispersion analysis allows us to 
reinforce our understanding of the underlying physics governing the 
early evolution of a cloud --- the precursor to the later nonlinear 
phase of evolution --- as well as provide a useful benchmark and 
theoretical standard to establish the overall accuracy of our numerical
code.  

Because we are concerned with the linear stages of gravitational 
instability, we will confine our focus of the numerical simulations in
this study to the evolution of the column density in model clouds. The 
detailed time-dependent behavior of other quantities, such as the 
velocity fields of the ions and neutrals, and the evolution and 
redistribution of mass in magnetic flux tubes will be described in a 
following paper.
\vspace{-2ex}
\subsubsection{Monochromatic perturbations}
We follow the evolution of clouds initially given perturbations 
with a single wavelength $\lambda$. The evolution of a model
cloud is initiated by imposing at time $t = 0$ an initial 
(normalized) column density profile of the form
\beqarr
\label{sigmainitstate}
\sign(x,y,0) = 1 + \delsigamp\cos(2 \pi x/\lambda),
\eeqarr
that is, a uniform background state with a perturbation with amplitude 
$\delsigamp$. In using this particular perturbation, we make use of the 
fact that, as mentioned in \S~3.1 above, the dispersion analysis 
indicates that the effects of linear disturbances are independent of the
angle of propagation $\theta$. Since it really doesn't matter for our 
purposes which direction of propagation we choose, we take 
$\theta = 0$ (parallel to the $x$-axis), which means that $\ky = 0$, and
therefore $\kx = \kz = 2 \pi/\lambda$. The initial velocity and magnetic
field are also perturbed in a way that is consistent with 
the system of equations (\ref{linconteq}) - (\ref{linBeq}) for the 
column density perturbation specified by equation 
(\ref{sigmainitstate}). Solving for the initial perturbations 
$\delta\vnx$, $\delta\Beq$, and $\delta\vny$ in terms of the given 
$\delta\sign$, $\kx$, and $\tgrow = i/\omega$ (which is a function of 
$\lambda$) from this set of equations yields
\beqarr
\label{vnxinitstate}
\delta\vnx(x,y,0) &=& - \frac{\lambda}{2 \pi \tgrow}
\delsigamp \sin(2 \pi x/\lambda), \\
\delta\Beq(x,y,0) &=& \frac{\lambda \Breftil}
{\lambda + 2 \pi \tgrow \tniitil \Breftil^{2}(1 + 2 \pi \Zi/\lambda)}
\delsigamp \cos(2 \pi x/\lambda)~,  
\eeqarr
and $\delta\vny(x,y,0) = 0$. 
By specifying the relation between the perturbed physical variables in 
this way, we are basically selecting the eigenvector of the perturbation
at the wavelength $\lambda$. Hence, the monochromatic perturbation 
excites a single eigenmode of a model cloud at $t=0$. 

Not surprisingly, when we initiate the time evolution in this fashion, 
the subsequent evolution of a model cloud is simply the continued growth
of the excited lone eigenmode. The left panel of Figure 5 displays the 
growth of the column density maxima for four different cloud models as a
function of time. Each has $\tniitil = 0.2$ and $\Pexttil = 0.1$. For 
all of these models, $\delsigamp = 0.02$ and $\lambda = 4 \pi$. The 
calculations were performed on an equally spaced mesh of 32 $\times$ 32
cells on a square computational region of size 
$L=4 \pi$ in each direction. Shown in that Figure (solid curves) is the
evolution of the maximum reduced column density 
$\delta\sign(t) = \sign(t) - 1$ for models with $\mui = 0.5$, 1, 2, and 
10. Also shown (dashed curves) is the growth of the column density 
perturbations that would occur as predicted by the linear relation for 
an unstable mode at the same location, 
$\delta \sign(t) = \delsigamp\exp(t/\tgrow)$. The values of $\tgrow$ 
used to plot the curves are those derived from the earlier dispersion 
analysis for $\lambda = 4 \pi$, $\tniitil=0.2$ and $\Pexttil=0.1$. From
the data in Figure $1d$ we find that 
$\tgrow =$ (20.3, 16.0, 4.34, 2.12) for $\mui =$ (0.5, 1.0, 2, 10). 
Comparing the numerical simulation results to the 
theoretical values predicted by the linear analysis indicates that they
are in excellent agreement and overlap during much of the early 
evolution of each model. In fact, significant deviation ($\gtrsim 1\%$)
between the simulation results and the linear predictions does not begin
to occur until $\delsign$ has grown to $\sim 0.2$, which is well beyond
what is generally considered the regime of linear growth. At later times
nonlinear effects are evident, and the column density grows more rapidly
and significantly exceeds that predicted by the linear analysis by the 
end of the period shown in Figure 5, when $\delsign = 1$.

The growth of the column density perturbations in the numerical 
simulations at early times is entirely consistent with our discussion 
of the physical processes acting in the dispersion analysis in \S~3.1. 
For instance, the linear analysis predicts that the onset of 
gravitational instability in the subcritical $\mui = 0.5$ model is due 
to the action of ambipolar diffusion (see Fig. $1d$). Moreover, using 
the stationary field limit approximation (eq. [\ref{tgrstationary}]) to
calculate $\tgrow$ for this model ($\lamt = 6.75$ for these parameters)
gives $\tgrow = 21.8$, which differs from the full dispersion analysis 
result quoted above by only $7\%$. The critical model $\mui = 1$ and 
supercritical model $\mui = 2$ are also heavily influenced by ambipolar
diffusion, and they would not be able to collapse in its absence, 
because the excited wavelength for both of these models is in the region
$\lamt < \lambda < \lamms$. Thus, these two models lie in the transition
region between diffusion-regulated and flux-frozen collapse for this 
particular wavelength, as noted in our discussion of these models in 
regard to Figure $1d$. Finally, the instability of the highly 
supercritical model $\mui = 10$ occurs essentially with freezing of 
magnetic field lines in the neutral matter. For this model, 
$\lamms = 6.93$ which is $< 4 \pi$. Using the flux-frozen approximate 
expression for the growth time given by equation (\ref{tgrfluxfreeze}) 
yields $\tgrow = 2.12$ for this model, which is exactly the same as the
full dispersion relation solution, and is also in very good agreement 
with the model simulation's early-time linear growth. 

The right panel of Figure 5 shows the growth of the column density
maxima in another four model clouds, with the same wavelength and
parameters except $\Pexttil = 10$. The development of these 
higher-$\Pexttil$ models is again seen to be in excellent agreement with
the dispersion analysis for the linear stage of evolution. They are also
in accordance with the values of $\tgrow$ predicted by the analytic 
relations (\ref{tgrstationary}) and (\ref{tgrfluxfreeze}) for the 
subcritical and supercritical models, respectively.

We conclude then, that our numerical code accurately reproduces the 
physics of gravitational instability in planar magnetic model clouds. 
\vspace{-2ex}
\subsubsection{White noise perturbations: the wavelength of 
minimum growth time $\lammin$ and its dependence on $\mui$}
In this subsection we consider cloud models that are given a spectrum of
random, small-amplitude perturbations in the physical variables at 
$t=0$. The spectrum we use is white noise, i.e, flat, so that there
is no preferred wavelength selected by the perturbations. However, we 
introduce damping so that wavelengths equal to twice the mesh spacing 
and smaller are negligible. The root-mean-square amplitude of the 
fluctuations in the initial state of a model cloud is 3\% of the uniform
background. The size of the computational region in each direction is 
$L=16 \pi$, and the number of cells in each direction is $N=128$. 

The white noise spectrum will excite an ensemble of eigenmodes with a 
wide range of wavelengths. The dominant evolutionary modes that will 
emerge from this ensemble of fluctuations and govern the subsequent 
evolution will be those grouped about the one with the minimum growth 
time. A cloud will then develop features with lengthscales comparable to
the wavelength with the minimum growth time, $\lammin$. That this 
intuitive notion is valid can be seen in Figure 6, which shows contour 
maps of the column density $\sign(x,y)$ in two model evolution 
simulations at the time for which the value of the maximum column 
density has grown to $\sigma_{\rm n, max} = 2$ (twice the initial 
background value) in each model. Both models again have $\tniitil =0.2$
and $\Pexttil=0.1$. The model displayed in the left panel is initially 
subcritical with $\mui = 0.5$, and the one in the right panel started as
a critical cloud with $\mui = 1$. The density contours cover the range 
0.8 to 1.6, in increments of 0.2. By the time that the contour snapshots
are taken, nonlinear effects have set in. This is evidenced by column 
densities that are relatively large compared to that of the initial 
background state, and the beginning of nonaxisymmetric gravitational 
fragmentation (which will form protostellar cores) due to the 
interaction of various eigenmodes. Despite this, both cloud models 
retain a significant imprint from the predecessor linear stage of 
evolution: there is a nearly uniform separation between core fragments
--- defining here a core as the enclosed higher column density regions 
with $\sign > 1.2$ --- and the fragmentation scale for these nascent 
cores is close to $\lammin$ in each model. For the model with 
$\mui = 0.5$, Figure 2 indicates that $\lammin = 14$, while for the 
cloud with $\mui = 1$, $\lammin = 24$. Examination of Figure 6 reveals 
that these scales correspond to the mean distance between the different
$\sign = 1.2$ contours for these models. Additionally, we note that the
behavior of the fragmentation scale in these models as a function of 
$\mui$ is also consistent with the linear analysis. Notably, the average
spacing between cores in the critical model evolution simulation is 
larger than that in the subcritical model, in line with the predicted 
dependence of $\lammin$ on $\mui$ (see Fig. 2). This results in there 
being fewer cores within the same size region. In fact, 
according to Table 1 (see, also, Fig.  $1d$), the maximal 
$\lammin (=52.8)$ for these values of $\tniitil$ and $\Pexttil$ occurs 
at $\mui = 1.13$. 

Further confirmation that $\lammin$ sets the characteristic lengthscales
in clouds can be seen in Figure 7, which presents the column density
plots of two more model clouds, with the same values of $\tniitil$
and $\Pexttil$ as in Figure 6. The model in the left panel of Figure
7 is supercritical and has $\mui = 2$. The linear analysis predicts
$\lammin = 23$ for this model, which is indeed close to the the 
distance between core-bounding contours in that Figure. There are also 
relatively fewer cores that have formed, similar to that seen in the 
critical model. The similarity in the {\em structural (lengthscale)} 
properties of the $\mui = 1$ and $\mui =2$ models is not surprising, as
they have nearly equal values for $\lammin$. However, the 
{\em dynamical} properties of the $\mui=1$ and $\mui=2$ models are 
quite different in the nonlinear regime, as shown by BC04. Specifically,
the maximum velocities in the supercritical models
become supersonic on scales that are well within the resolving power of
modern observations. We defer the detailed discussion of the dynamical 
and related nonlinear evolution of cloud models to a study that will 
follow this paper, and to that already presented in BC04. Finally, the 
right panel of Figure 7 shows the column density 
contours for a highly supercritical model with $\mui = 10$. For this 
model, $\lammin = 13$. This is also in agreement with the mean distance 
between cores in the contour plot. The lengthscales and number of 
cores in the highly supercritical model are akin to that seen in the 
subcritical $\mui = 0.5$ model. This is no coincidence, since these two
models have similar values of $\lammin$. Although a somewhat 
counterintuitive result, this is because there is also a Jeans-like 
growth of density perturbations in the subcritical models, but it 
occurs on the much longer ambipolar diffusion timescale as neutrals 
drift past near-stationary plasma and magnetic field. 

There are also some differences in the models in Figures 6 and 7 that 
become more pronounced with increasing initial mass-to-flux ratio: the 
cores become more nonaxisymmetric with greater values of $\mui$, showing
enhanced elongation along a single axis. This second-order effect --- 
an amplification of nonaxisymmetry due to the nonlinear interaction of 
certain eigenmodes for the more supercritical models --- will be a topic
of investigation in a following paper. And, as mentioned above, the 
dynamical properties of the model clouds, such as their velocity fields,
are also markedly different, characteristically having greater infall 
velocities over larger scales with increasing $\mui$. This too 
will be studied further in a future paper. Despite these differences, we
conclude from our representative evolution simulations that the 
gravitational fragmentation scale of a model cloud with an initial 
spectrum of perturbations is effectively the value of $\lammin$
for that particular model. The fragmentation scale has a dependence on 
$\mui$ in agreement with the result of the linear analysis, especially 
with the prediction that near the critical value there is a dramatic 
increase in $\lammin$. Because of this resonant-like behavior,
{\em clouds that are close to critical will have significantly larger
size scales than in models that are more highly supercritical or 
more highly subcritical.} 
\vspace{-2ex}
\section{Summary}

We have presented the formulation of physical models 
that we will use to study the nonaxisymmetric formation and 
self-gravitational collapse of protostellar cores in magnetic 
interstellar molecular clouds. Model clouds are partially ionized
isothermal thin planar sheets with finite half-thickness $Z(x,y)$.

The system of equations that are used to govern the time evolution
of model clouds contain four fundamental parameters, all defined in 
\S~2.3. The first is the dimensionless background magnetic field 
strength of a cloud, $\Breftil$; the inverse of this parameter is the 
initial mass-to-magnetic flux ratio in units of the critical value for 
gravitational collapse, $\mui$. Clouds with $\Breftil > 1$ ($\mui < 1$)
are subcritical and are initially magnetically supported, and cannot 
collapse in the absence of ambipolar diffusion. Models with 
$\Breftil < 1$ ($\mui > 1$) are supercritical from the outset and unable
to support themselves against their own self-gravity. The next parameter
is the normalized neutral-ion collisional momentum-exchange time 
$\tniitil$. It is a measure of the efficiency of ambipolar diffusion in
model clouds: $\tniitil \rightarrow 0$ corresponds to very effective 
momentum transfer and collisional coupling of neutrals and ions, 
resulting in freezing of magnetic flux in the neutral matter. In the 
opposite limit of $\tniitil \gg 1$, collisions are so infrequent that 
the neutrals are essentially decoupled from the plasma and magnetic 
field, and magnetic forces contribute negligibly to the support and 
evolution of a model cloud. $\Pexttil$ is the ratio of the external 
pressure to the vertical self-gravitational stress in the initial 
uniform reference state of a model cloud. The final parameter is the 
exponent $k$ in the power-law expression used to calculate the ion 
density as a function of the neutral density. 

We also investigated the linear stability of model clouds, and how 
the growth times $\tgrow$ and critical wavelengths of the 
gravitationally unstable modes depend on $\mui$, $\tniitil$, and 
$\Pexttil$. Analytic expressions for $\tgrow$ and the critical 
wavelengths were derived for the limits of weak 
and strong ambipolar diffusion. These expressions (eqs. 
[\ref{tgrfluxfreeze}] and [\ref{tgrstationary}]) agreed well with the 
full dispersion results in these limits. 

Models with frozen-in magnetic flux ($\tniitil = 0$) are 
gravitationally unstable when they are supercritical and the 
wavelength of a perturbation exceeds the magnetosound critical 
wavelength $\lamms$. For finite thickness clouds, magnetic pressure 
contributes substantially to setting the value of $\lamms$ and $\tgrow$
when $\mui \sim 1$. Ambipolar diffusion ($\tniitil > 0$) allows 
clouds to be unstable even when subcritical, so long as the 
perturbation's wavelength is greater than the thermal critical 
wavelength $\lamt$ ($\lamt \leq \lamms$, see eqs. [\ref{lammsdefeq}] and
[\ref{lamtdefeq}]). The instability modes of clouds with $\Pexttil > 1$
behave qualitatively the same way as for those with $\Pexttil \leq 1$, 
except that quantitatively their critical wavelengths and growth times 
are reduced, due to the effective retarding sound speed being decreased
at higher external pressures. The dispersion analysis also revealed how 
the wavelength with the minimum growth time, $\lammin$, behaves as
a function of $\mui$. For models with $\tniitil > 0$, $\lammin$ has a 
resonance at a value of $\mui$ that is usually just slightly greater 
than the critical value $\mui = 1$. Because of this resonance, a model 
cloud with $\mui \sim 1$ will have a $\lammin$ that is significantly 
greater when compared to that of models that are much more sub- or 
supercritical. 

In addition to providing basic insight to the physics of gravitational
instability in partially ionized media, the linear analysis was also
used as a theoretical standard to test the accuracy of our numerical 
method of solution of the full set of nonlinear equations (eqs. 
[\ref{massconteqb}] - [\ref{Byeqb}]) that govern the time evolution of a
model cloud. The early-time evolution of several cloud simulations given
an initial monochromatic perturbation of wavelength $\lambda$ was 
compared to the predicted value of the growth timescale from the 
dispersion analysis. The temporal growth of the column density maxima in
each model simulation was in excellent agreement with the 
theoretically predicted values of $\tgrow$, thus establishing 
the accuracy of our non-ideal magnetohydrodynamic computational code.

Finally, we presented a suite of model simulations that had their 
evolution initiated by fluctuations with a spectrum of wavelengths,
following their evolution a little beyond the linear growth phase. The 
characteristic fragmentation scale that developed in these models tends 
to correspond to the wavelength with the minimum growth time $\lammin$ 
of the initial state. The resonance behavior of $\lammin$ for clouds 
with the parameter $\mui$ near the critical value --- as predicted by 
the linear analysis --- was also in evidence in the numerical 
simulations. Those having $\mui \sim 1$ had a much greater mean spacing
($\approx \lammin$) between cores than in models with $\mui \ll 1$ and 
$\mui \gg 1$. Because of this, the total number of cores that developed
in the models near the resonance was less than in those that were 
farther away. This sensitive dependence of $\lammin$ about $\mui \sim 1$
has important implications for core and star formation, since 
observations currently indicate that mass-to-flux ratios in molecular 
clouds generally lie in the range $0.5 \lesssim \mui \lesssim 2$ 
(Crutcher 2004).

In forthcoming studies we will explore further the formation and 
nonaxisymmetric collapse of protostellar cores as a function of the 
fundamental model parameters, building on the formulation and analysis 
presented in this paper. The properties of dynamically infalling cores 
such as spatial density and velocity maps, core shapes, magnetic field 
strengths, and other quantities of interest, will be investigated in 
detail. In doing so, we will be providing a physically consistent model 
with testable, quantitative predictions that may be used to interpret 
and perhaps guide observations of star formation in magnetic
interstellar molecular clouds. 

\acknowledgments{GC was supported by the New York Center for the
Origins of Life (NSCORT) and the Department of Physics, Applied
Physics, and Astronomy at Rensselaer Polytechnic Institute, under
NASA Grant NAG5-7589. SB was supported by a grant from the Natural
Sciences and Engineering Research Council of Canada. Helpful comments
from an anonymous referee are gratefully acknowledged.}
\clearpage
\begin{table}[hb]
\begin{center}
Table 1 \\
{\sc Peak ratio of wavelength with minimum growth time ($\lammin$) to
wavelength of maximum gravitational instability ($\lamtm$)} \\ 
\bigskip
\begin{tabular}{llccc}
\hline
\hline
\\
\mbox{\hspace{1em}} & $\tniitil$ & $~~\mui~~$ & 
$~~\left(\frac{\lammin}{\lamtm}\right)_{\rm peak}~~$ 
\\ \\
\hline
\\
$\Pexttil = 0.1$: & & & & \\
& 0.04 & $~~1.02~~$ & $~~21.1~~$ \\           
& 0.1 & $~~1.06~~$ & $~~8.37~~$ \\
& 0.2 & $~~1.13~~$ & $~~4.20~~$ \\
& 1 & $~~1.64~~$ & $~~1.42~~$ \\
& 10 & $~~3.50~~$ & $~~1.04~~$ \\
$\Pexttil = 10$: & & & & \\
& 0.04 & $~~1.06~~$ & $~~6.04~~$ \\
& 0.1 & $~~1.11~~$ & $~~2.89~~$ \\
& 0.2 & $~~1.23~~$ & $~~1.90~~$ \\
& 1 & $~~1.87~~$ & $~~1.17~~$ \\
& 10 & $~~3.31~~$ & $~~1.01~~$ \\
\\
\hline
\end{tabular}
\end{center}
\end{table}

\clearpage
\begin{figure}
\epsscale{0.9}
\plotone{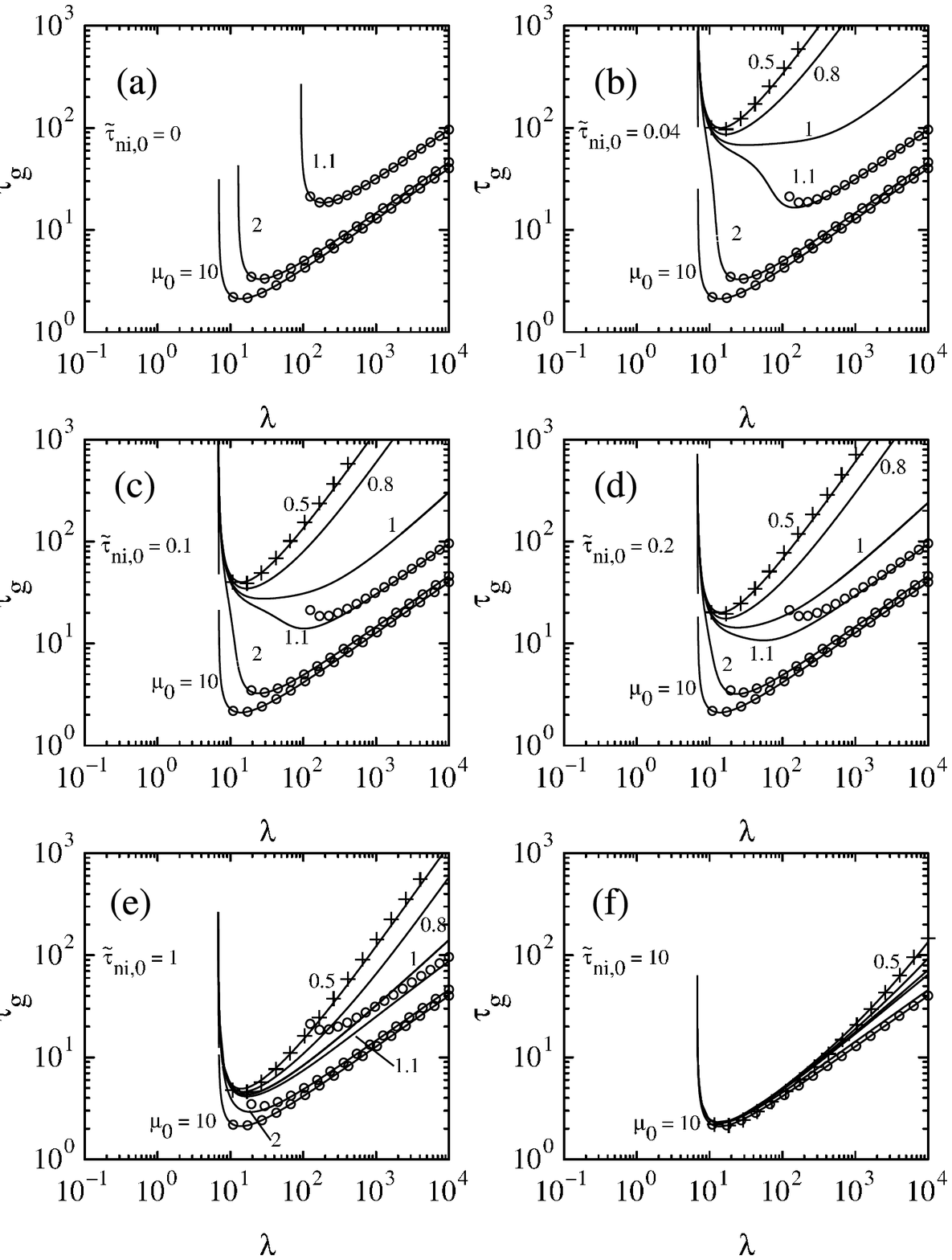}
\caption{Instability growth times for various model clouds as a function
of wavelength, for models with ({\it a}) $\tniitil = 0$, 
({\it b}) $\tniitil = 0.04$, 
({\it c}) $\tniitil = 0.1$, ({\it d}) $\tniitil = 0.2$, ({\it e}) 
$\tniitil = 1$, and ({\it f}) $\tniitil = 10$, respectively. In
each panel are shown the timescale curves for models with mass-to-flux 
ratios $\mui = 0.5$, 0.8, 1, 1.1, 2, and 10 (labeled). Each model has 
$\Pexttil = 0.1$. Also shown are the results of
the approximate analytical solutions in the limit of flux-freezing
(open circles), and stationary magnetic field lines with ambipolar 
diffusion (crosses), given by eqs. (\ref{tgrfluxfreeze}) and
(\ref{tgrstationary}), respectively.}
\end{figure}

\begin{figure}
\epsscale{0.4}
\plotone{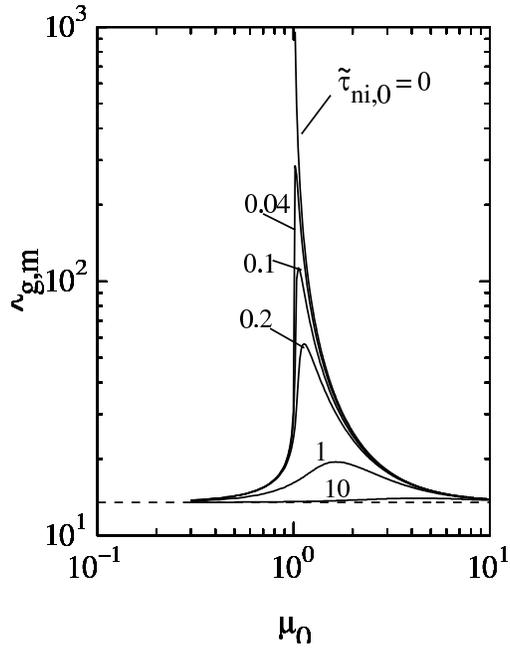}
\caption{Wavelength with the minimum growth time (= maximum growth
rate) as a function of initial mass-to-flux ratio. The displayed curves 
are for models with $\tniitil = 0$, 0.04, 0.1, 0.2,
1, and 10, respectively. All models have $\Pexttil = 0.1$.
The dashed curve is the value of $\lamtm$ for these models.}
\end{figure}

\begin{figure}
\epsscale{0.95}
\plotone{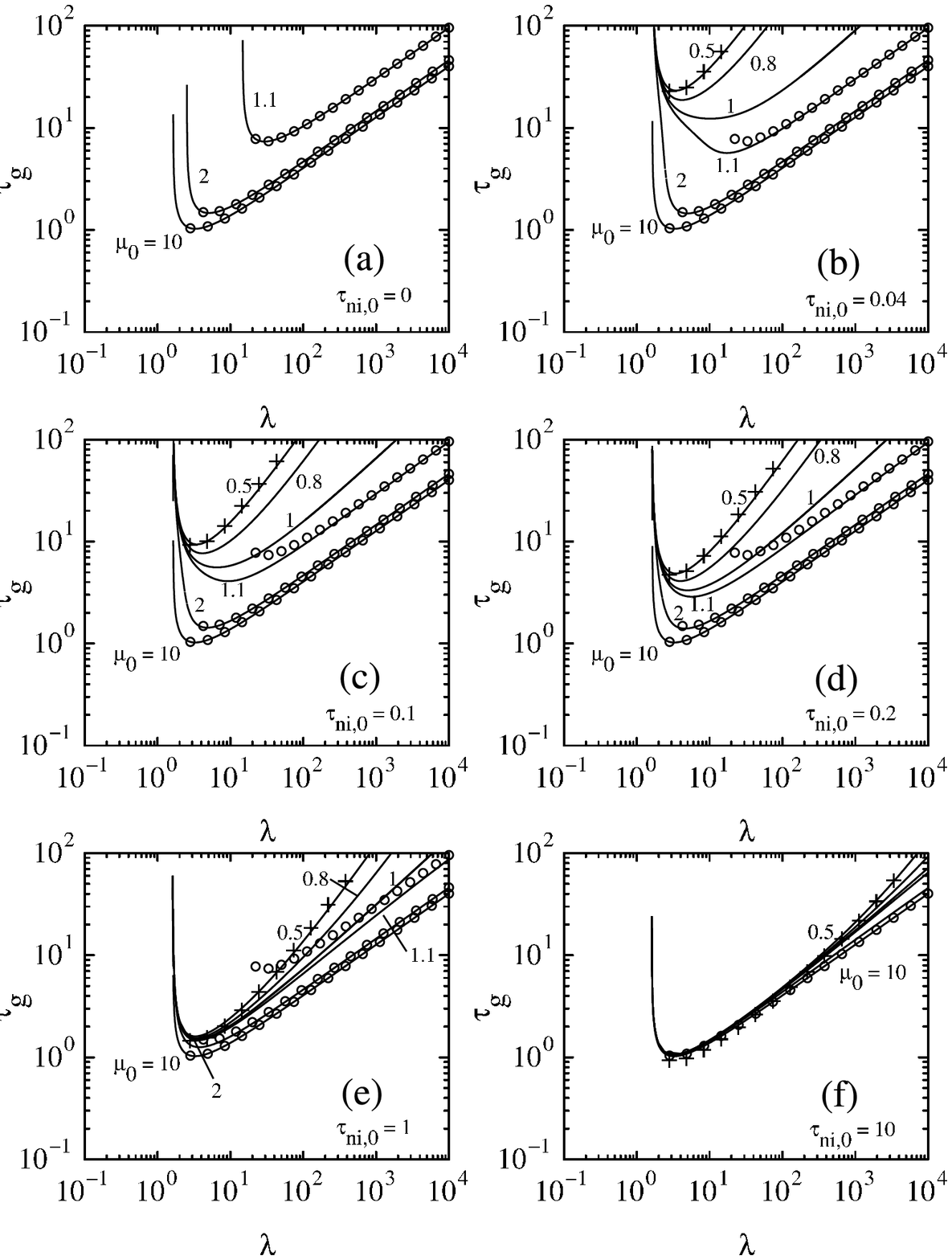}
\caption{Same as in Fig. 1, but with $\Pexttil = 10$.}
\end{figure}

\begin{figure}
\epsscale{0.4}
\plotone{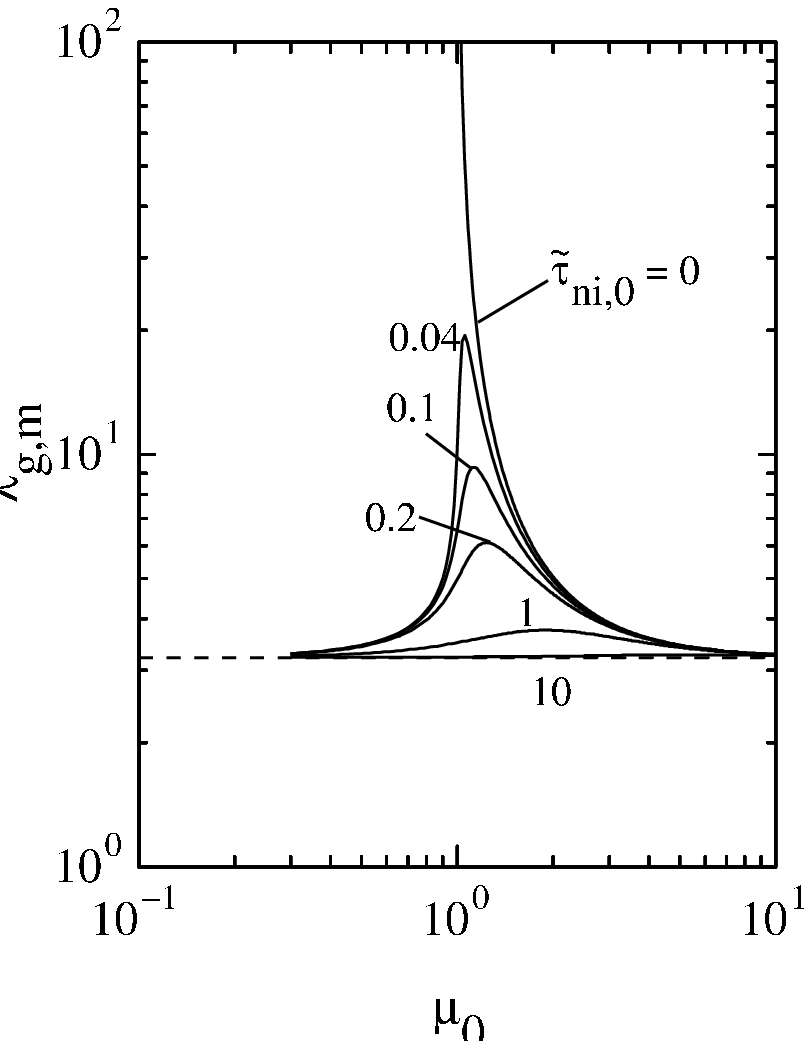}
\caption{Same as in Fig. 2, but with $\Pexttil = 10$.}
\end{figure}

\begin{figure}
\epsscale{1.1}
\plottwo{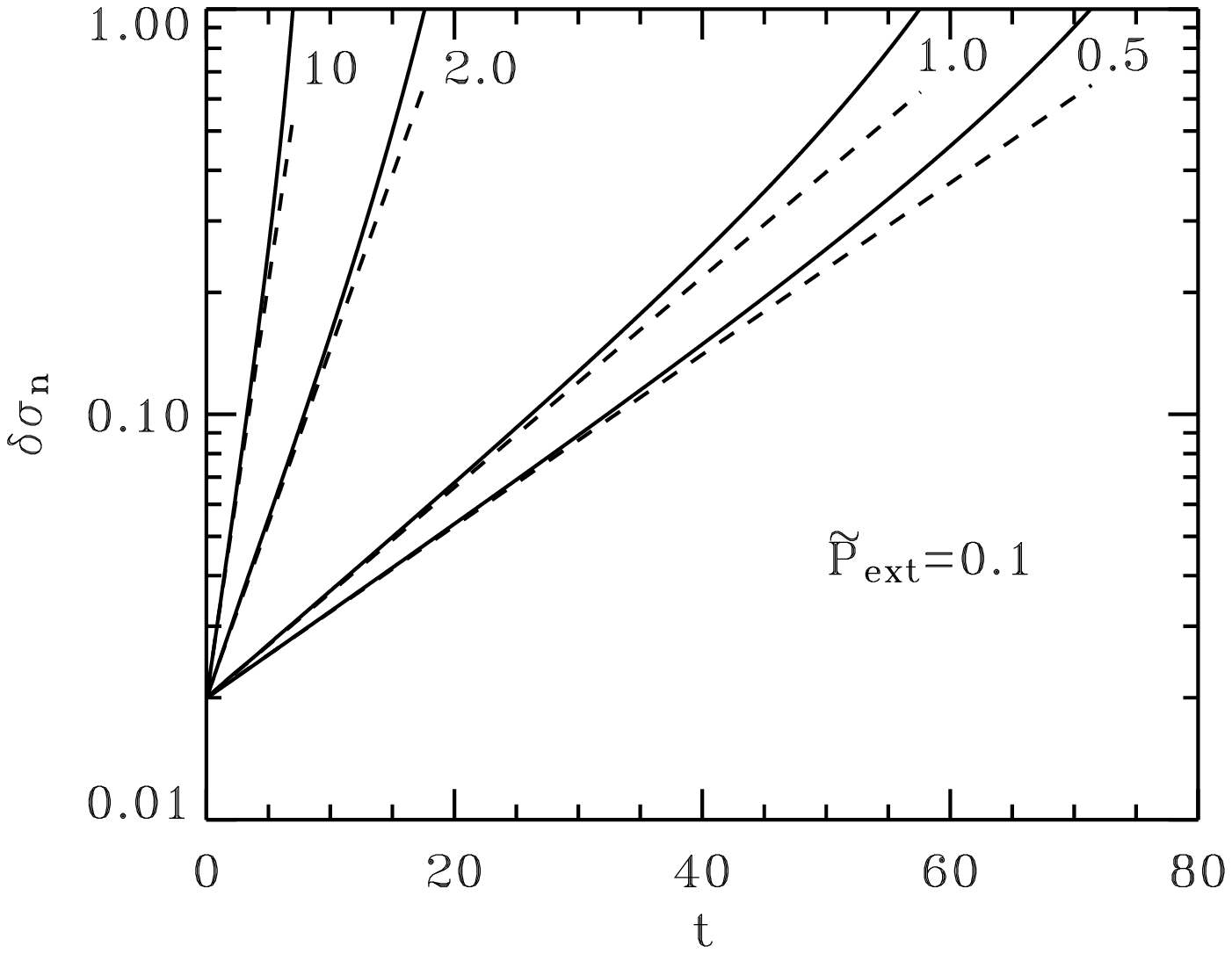}{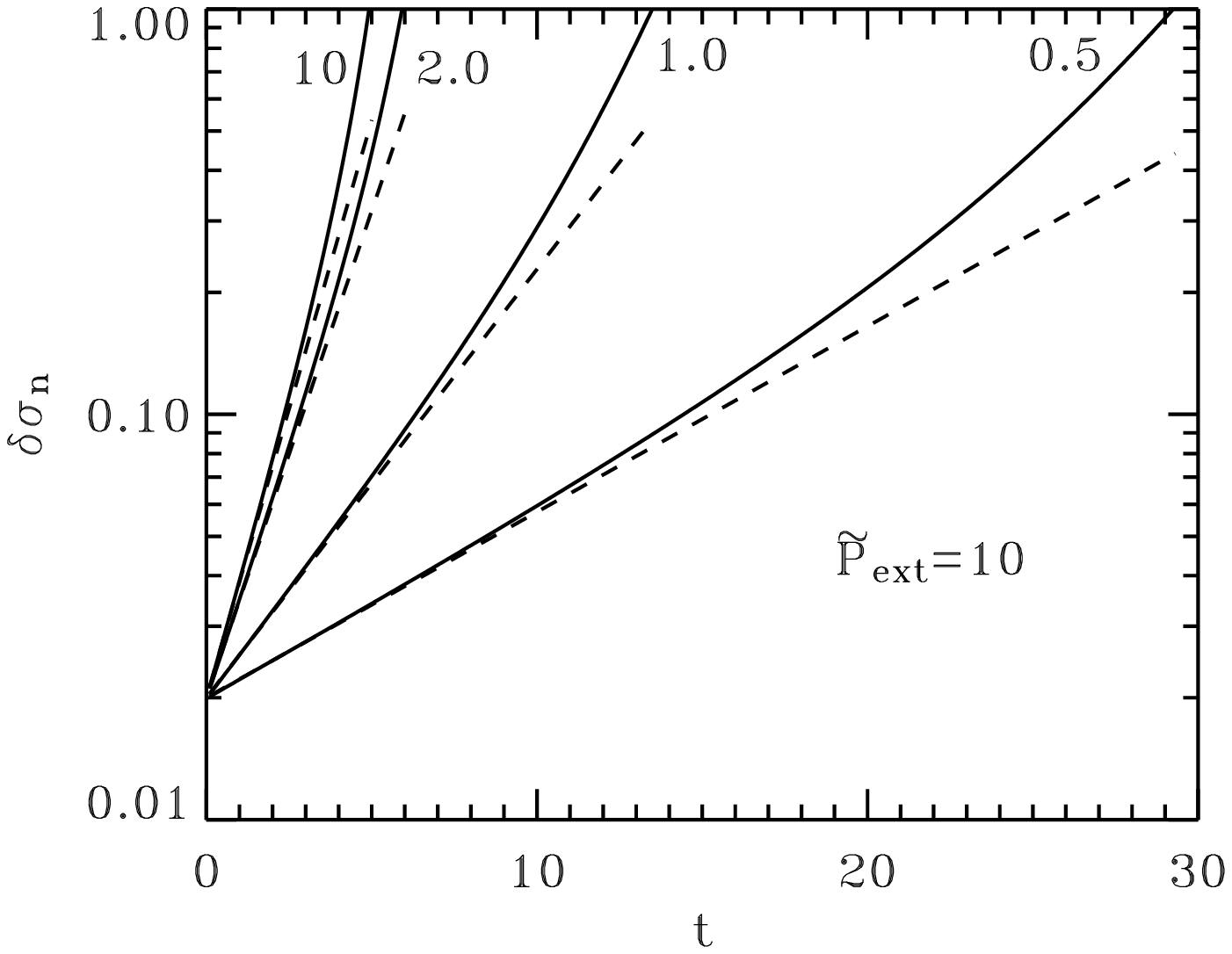}
\caption{Time evolution of column density maxima for numerical 
simulations of various cloud models given an initial
monochromatic perturbation of wavelength $\lambda = 4 \pi$. 
The amplitude of the initial perturbation of the column density is 
$\delsigamp = 0.02$. Each model has $\tniitil = 0.2$, and is labeled by
the value of its mass-to-flux ratio $\mui$, which is $0.5$, $1$, $2$, 
and $10$, respectively. Left panel: models with $\Pexttil = 0.1$. Right:
models with $\Pexttil = 10$. The solid lines show the peak density in 
each model simulation. The dashed lines are the theoretical values given
by the relation $\delta \sign(t) = \delsigamp\exp(t/\tgrow)$, where 
$\tgrow$ is the growth time for each model predicted from the linear 
analysis.} 
\end{figure}

\begin{figure}
\epsscale{1.1}
\plottwo{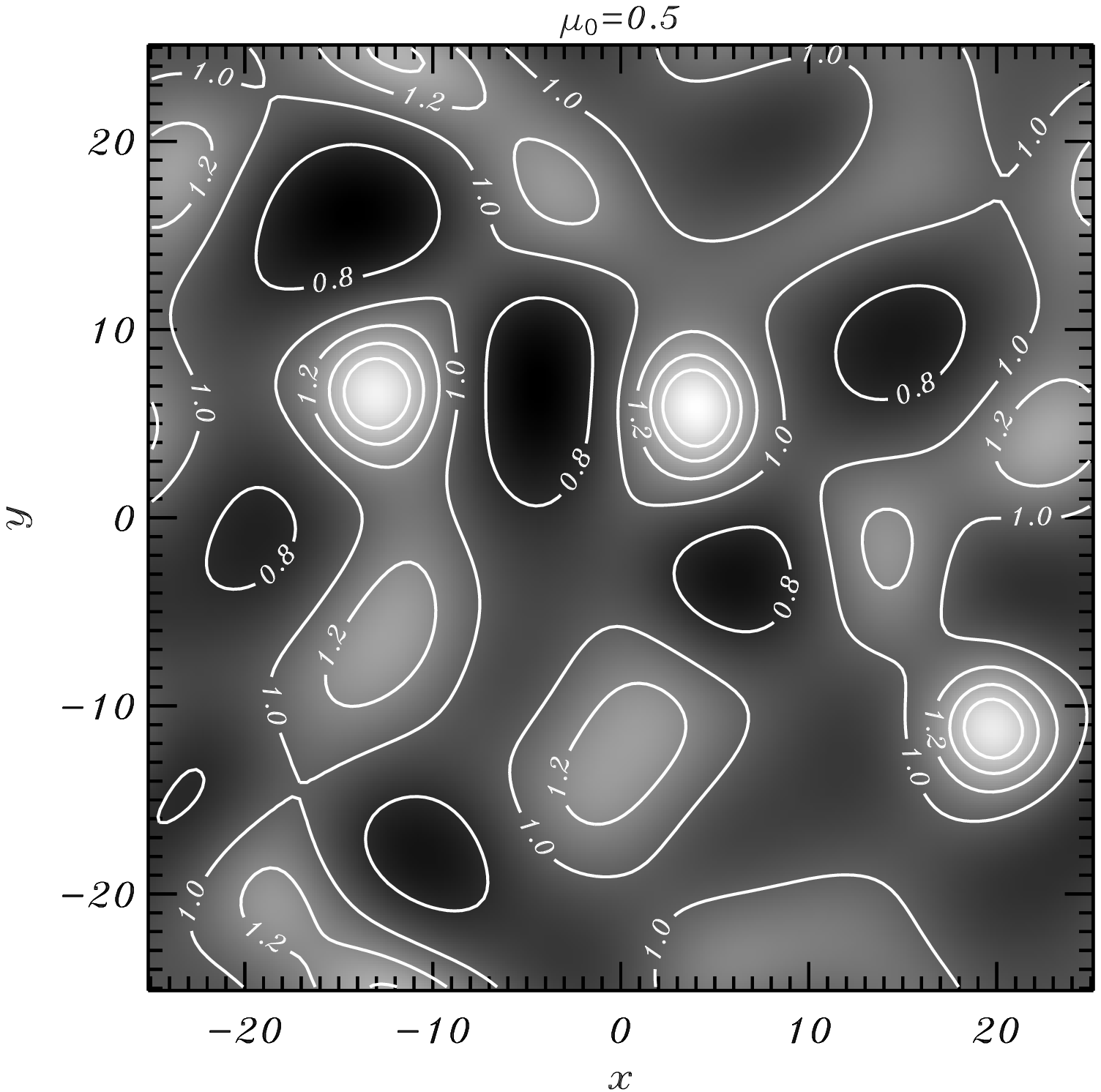}{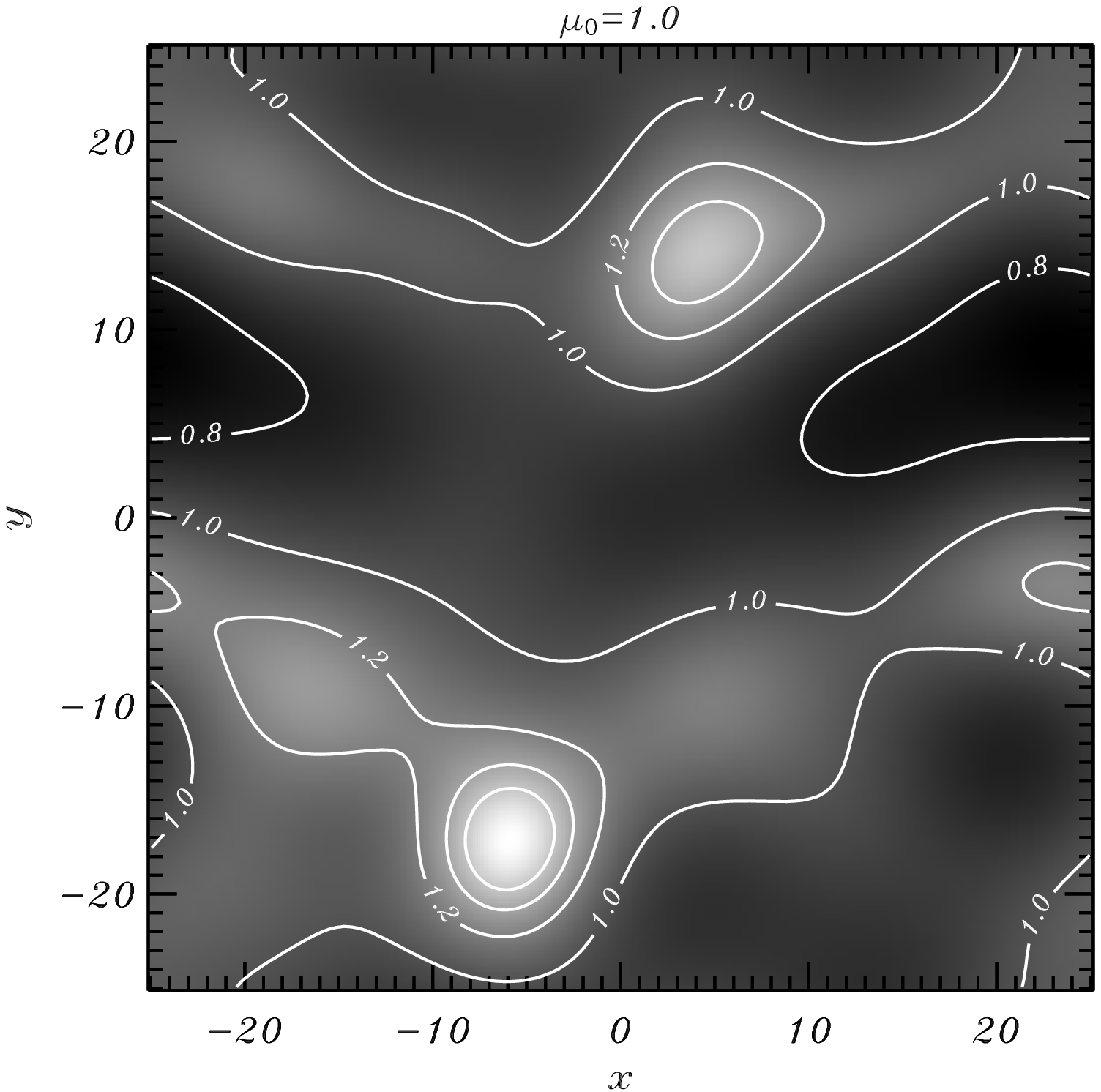}
\caption{Column density contours of two model clouds that initially
had random small-amplitude fluctuations in physical variables superposed
on a uniform background state. The contours are overlaid on a grayscale
image of the logarithm of the column density. At the time shown, the 
peak column density in each model is twice the value of the background 
state ($\sigma_{\rm n,max} = 2$). Each model has $\tniitil = 0.2$ and 
$\Pexttil = 0.1$. The left model has $\mui = 0.5$, and the right one has
the critical value $\mui = 1$. The density contours range from 0.8 to 
1.6, in steps of 0.2.}
\end{figure}

\begin{figure}
\epsscale{1.1}
\plottwo{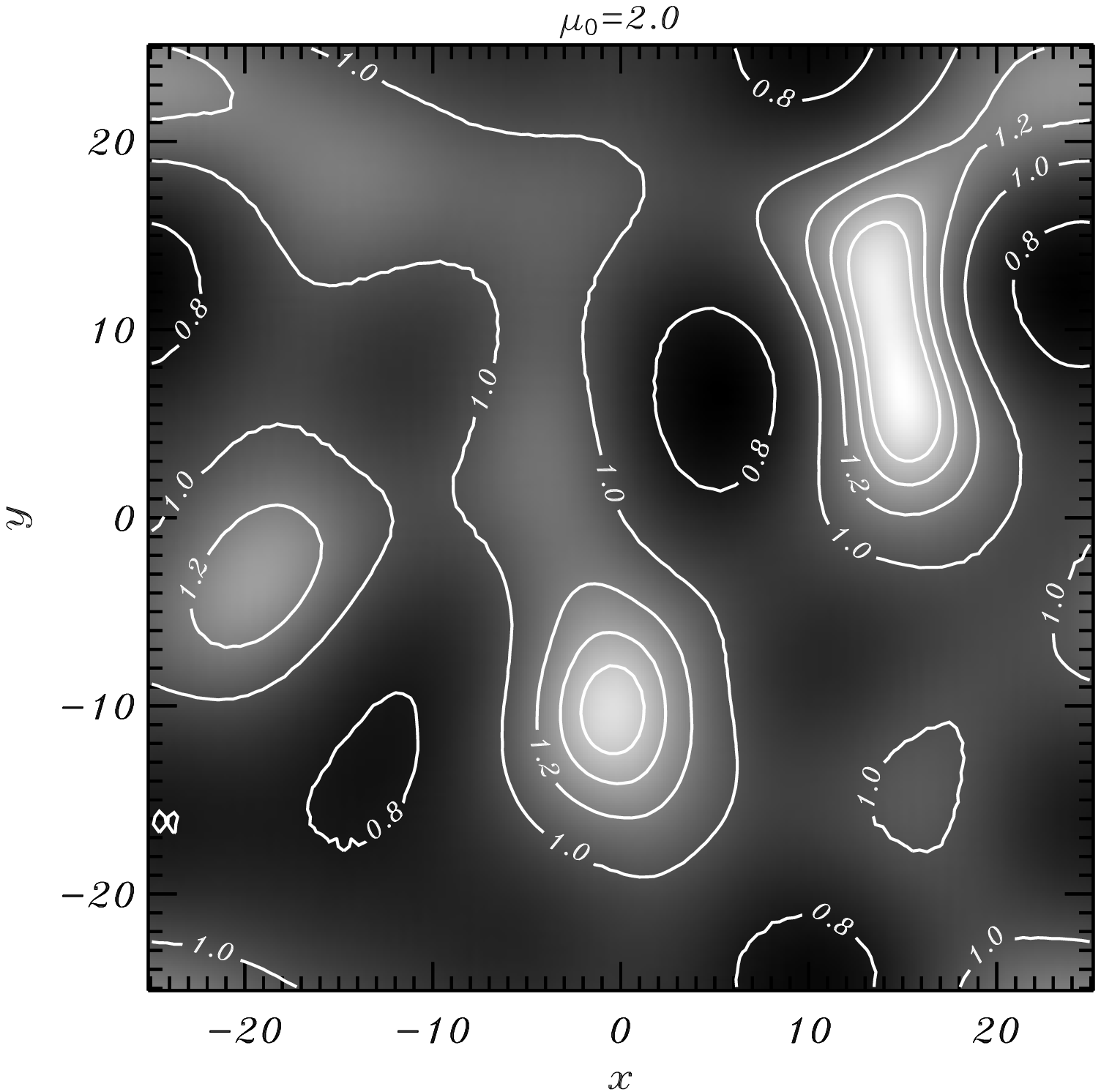}{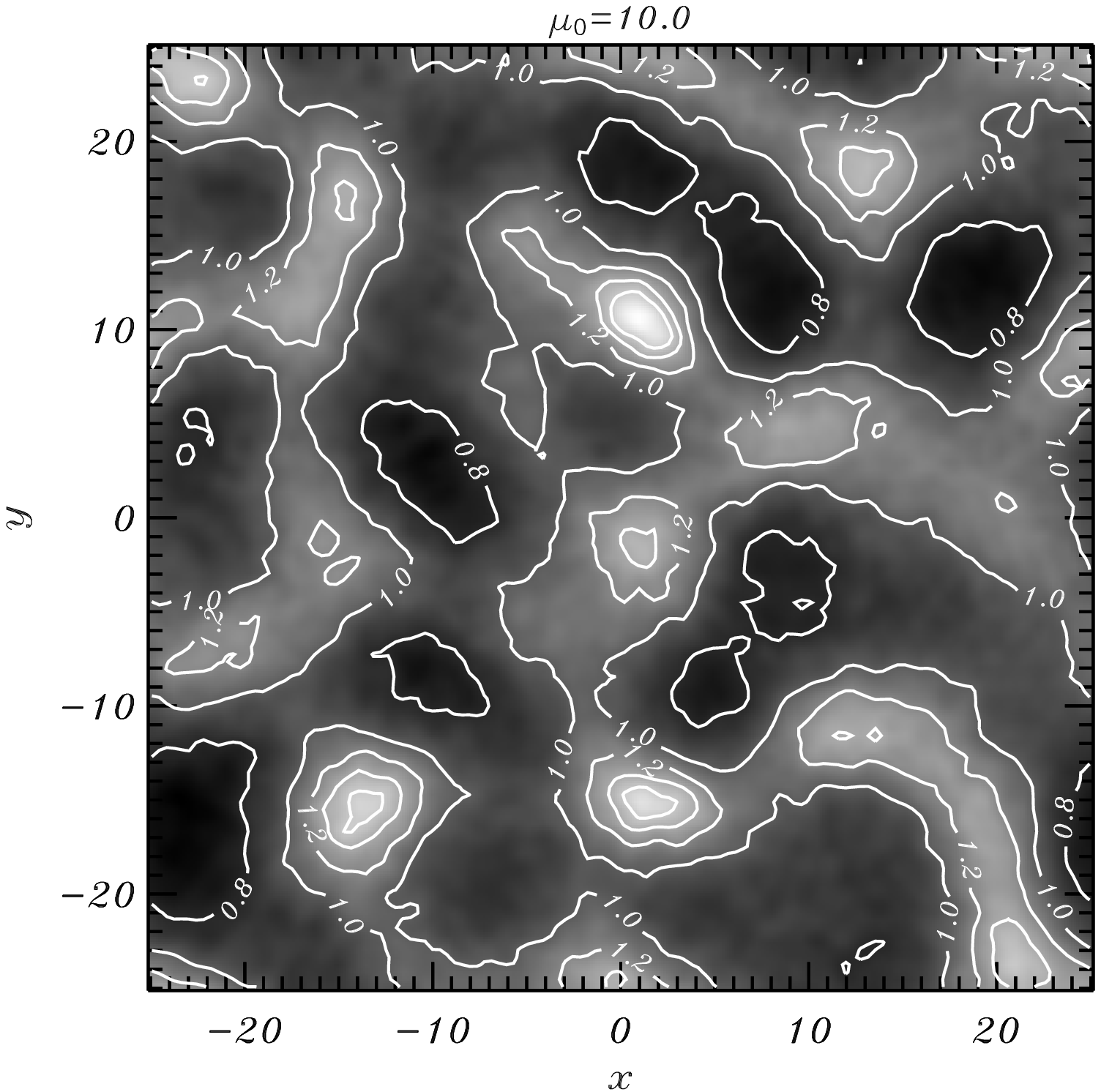}
\caption{Same as in Fig. 6, but for supercritical models with 
$\mui = 2$ (left) and $\mui = 10$ (right).}
\end{figure}


\clearpage

\begin{thebibliography}

\bibitem[]{AS00}Allen, A., \& Shu, F. H. 2000, \apj, 536, 368
\bibitem[]{A2001}Andr\'{e}, P., Motte, F., \& Belloche, A. 2001, in
From Darkness to Light: Origin and Evolution of Young Stellar Clusters,
ed. T. Montmerle \& P. Andr\'{e} (San Francisco: ASP), 209
\bibitem[]{Ba00}Bacmann, A., Andr\'{e}, P., Puget, J.-L., Abergel, A.,
Bontemps, S., \& Ward-Thompson, D. 2000, \aap, 361, 555 
\bibitem[]{B97}Basu, S. 1997, \apj, 485, 240 
\bibitem[]{B00}\ul. 2000, \apj, 540, L103
\bibitem[]{BC04}Basu, S., \& Ciolek, G. E. 2004, \apj, 607, L39 (BC04)
\bibitem[]{BM94}Basu, S., \& Mouschovias, T. Ch. 1994, \apj, 432, 720 (BM94)
\bibitem[]{BM95a}\ul. 1995a, \apj, 452, 386
\bibitem[]{BM95b}\ul. 1995b, \apj, 453, 271
\bibitem[]{BF92}Byron, F. W., Jr., \& Fuller, R. W. 1992, Mathematical
Methods of Classical and Quantum Physics (Mineola: Dover)
\bibitem[]{Ca02}Caselli, P., Walmsley, C. M., Zucconi, A., Tafalla, M.,
Dore, L., \& Myers, P. C. 2002, \apj, 565, 331
\bibitem[]{CF53}Chandrasekhar, S., \& Fermi, E. 1953, \apj, 118, 113
\bibitem[]{CB00}Ciolek, G. E., \& Basu, S. 2000, \apj, 529, 925 (CB00)
\bibitem[]{CM93}Ciolek, G. E., \& Mouschovias, T. Ch. 1993, \apj, 418, 774 (CM93)
\bibitem[]{CM94}\ul. 1994, \apj, 425, 142
\bibitem[]{CM95}\ul. 1995, \apj, 454, 194
\bibitem[]{CM96}\ul. 1996, \apj, 468, 749
\bibitem[]{CM98}\ul. 1998, \apj, 504, 280
\bibitem[]{CK98}Ciolek, G. E., \& K\"{o}nigl, A. 1998, \apj, 504, 257
\bibitem[]{CCK98}Contopoulos, I., Ciolek, G. E., \& K\"{o}nigl, A. 1998,
\apj, 504, 247
\bibitem[]{Cra04}Crapsi, A., Caselli, P., Walmsley, C. M., Tafalla, M.,
Lee, C. W., Bourke, T. L., \& Myers, P. C. 2004, \aap, 420, 957
\bibitem[]{Cr04}Crutcher, R. M. 2004, in The Magnetized Interstellar 
Medium, ed. B. Uyaniker, W. Reich, \& R. Wielebinski (Katlenburg-Lindau:
Copernicus GmbH), 123
\bibitem[]{Cretal04}Crutcher, R. M., Nutter, D. J., Ward-Thompson, D. 
W., \& Kirk, J. M. 2004, \apj, 600, 279
\bibitem[]{CT01}Crutcher, R. M., \& Troland, T. H. 2000, \apj, 537, L139
\bibitem[]{Cu04}Curran, R. L., Chrysostomou, A., Collett, J. L., 
Jenness, T., \& Aitken, D. K. 2004, \aap, 421, 195
\bibitem[]{FM93}Fiedler, R. A., \& Mouschovias, T. Ch. 1993, \apj, 415, 
680 
\bibitem[]{Fu01}Fujiyoshi, T., Smith, C. H., Wright, C. M., Moore,
T. J. T., Aitken, D. K., \& Roche, P. F. 2001, \mnras, 327, 233
\bibitem[]{IZ00}Indebetouw, R. M., \& Zweibel, E. G. 2000, \apj, 532, 
361
\bibitem[]{J02}Jones, C. E., \& Basu, S. 2002, \apj, 569, 280
\bibitem[]{J01}Jones, C. E., Basu, S., \& Dubinski, J. 2001, \apj, 551, 
387
\bibitem[]{K04}Kerton, C. R., Brunt, C. M., Jones, C. E., \& Basu, S.
2003, \aap, 411, 149
\bibitem[]{KNMH87}Kiguchi, M., Narita, S., Miyama, S., \& Hayashi, C.
1987, \apj, 317, 830
\bibitem[]{KB03} Kudoh, T., \& Basu, S. 2003, \apj, 595, 842
\bibitem[]{KB06}\ul. 2006, \apj, 642, 270
\bibitem[]{La01}Lai, S.-P., Crutcher, R. M., Girart, J. M., \& Rao, R. 
2001, \apj, 561, 864
\bibitem[]{La02}\ul. 2002, \apj, 566, 925
\bibitem[]{La78} Langer, W. D. 1978, \apj, 225, 95
\bibitem[]{LN04}Li, Z.-Y., \& Nakamura, F. 2004, \apj, 609, L83
\bibitem[]{LS97a}Li, Z.-Y., \& Shu, F. H. 1997a, \apj, 475, 237
\bibitem[]{LS97b}\ul. 1997b, 475, 251
\bibitem[]{MM73}McDaniel, E. W., \& Mason, E. A. 1973, The Mobility
and Diffusion of Ions and Gases (New York: Wiley)
\bibitem[]{M91}Morton, S. A. 1991, Ph.D. thesis, University of Illinois,
Urbana-Champaign
\bibitem[]{MMC94}Morton, S. A., Mouschovias, T. Ch., \& Ciolek, G. E.
1994, \apj, 421, 561
\bibitem[]{Mo99}Motte, F., Andr\'{e}, P., \& Neri, R. 1998, \aap, 336, 
150
\bibitem[]{M76}Mouschovias, T. Ch. 1976, \apj, 207, 141
\bibitem[]{M77}\ul. 1977, \apj, 211, 147
\bibitem[]{M78}\ul. 1978, in Protostars and Planets, ed. T. Gehrels
(Tucson: U. Arizona Press), 209
\bibitem[]{M79}\ul. 1979, \apj, 228, 475
\bibitem[]{MC99}Mouschovias, T. Ch., \& Ciolek, G. E. 1999, in The
Origin of Stars and Planetary Systems, ed. C. J. Lada \& N. D. Kylafis
(Dordrecht: Kluwer), 305
\bibitem[]{NL05}Nakamura, F., \& Li, Z.-Y. 2005, \apj, 631, 411
\bibitem[]{NN78}Nakano, T., \& Nakamura, T. 1978, \pasj, 30, 671
\bibitem[]{NHM84}Narita, S., Hayashi, C., \& Miyama, S. 1984, Prog.
Theo. Phys., 72, 1118
\bibitem[]{PM04}Pereyra, A., \& Magalh\~{a}es, A. M. 2004, \apj, 603, 
584
\bibitem[]{PTVF}Press, W. H., Teukolsky, S. A., Vetterling, W. T.,
\& Flannery, B. P. 1996, Numerical Recipes in Fortran 77: The Art
of Scientific Computing (Vol. 1 of Fortran Numerical Recipes), 2nd. Ed.
(New York: Cambridge)
\bibitem[]{Sc91}Schiesser, W. E. 1991, The Numerical Method of Lines:
Method of Integration of Partial Differential Equations (San Diego:
Academic)
\bibitem[]{Sch98}Schleuning, D. A. 1998, \apj, 493, 811
\bibitem[]{Sch01}Schleuning, D. A., Vaillancourt, J. E., Hildebrand, R.
H., Dowell, C. D., Novak, G., Dotson, J. L., \& Davidson, J. A. 2000,
\apj, 535, 913
\bibitem[]{Sp78}Spitzer, L., Jr. 1978, Physical Processes in the
Interstellar Medium (New York: Wiley-Interscience)
\bibitem[]{Ta99}Tafalla, M., Mardones, D., Myers, P. C., Caselli, P.,
Bachiller, R., \& Benson, P. J. 1998, \apj, 504, 900
\bibitem[]{TM05a}Tassis, K., \& Mouschovias, T. Ch. 2005a, \apj, 618, 769,
\bibitem[]{TM05b}\ul. 2005b, \apj, 618, 783
\bibitem[]{vL79}van Leer, B. 1979, J. Comput. Phys., 32, 101
\bibitem[]{Wi99}Williams, J. P., Myers, P. C., Wilner, D. J., \& Di
Francesco, J. 1999, \apj, 513, L61
\bibitem[]{Wy76}Wyld, H. W. 1976, Mathematical Methods for Physics
(Reading: Benjamin)
\bibitem[]{Zu01}Zucconi, A., Walmsley, C. M., \& Galli, D. 2001, \aap, 
376, 650
\bibitem[]{Z98}Zweibel, E. G. 1998, \apj, 499, 746
\end{thebibliography}
\end{document}